**Climate-driven trends in the streamflow records of a reference hydrologic network in Southern Spain**

P. Yeste [(1)], J. Dorador[(2)], W. Martin-Rosales[(3)] , E. Molero [(1)], M. J. Esteban-Parra[(4)], and F. J. Rueda[(5)(6)]

(1) Dept. Urban Design and Spatial Planning, University of Granada, Spain

(2) Dept. Stratigraphy and Palaeontology, University of Granada, Spain

(3) Dept. Geodynamic, University of Granada, Spain

(4) Dept. Applied Physics, University of Granada, Spain

(5) Water Research Institute, University of Granada, Spain

(6) Dept. Civil Engineering, University of Granada, Spain

**Corresponding author:**

María Jesús Esteban-Parra (esteban@ugr.es; (+34)958240021)

**Abstract**

Monthly streamflow records from a set of gauging stations, selected to form a reference hydrologic network, are analyzed together with precipitation and temperature data to establish whether the streamflows in the Guadalquivir River Basin have experienced changes during the last half of the XXth century that can be attributed to hydrological forcing. The observed streamflows in the reference network have undergone generalized and significant decreases both at seasonal and annual scales during the study period. Annual rainfall, though, did not experienced statistically significant changes. The observed trends in streamflows can be attributed to either land-use changes, or to the statistically significant changes exhibited both by yearly potential evapotranspiration values and by the seasonal distribution of precipitation. In the attribution work conducted using both data-based and simulation-based methods, the intra-annual redistribution of precipitation is shown to be the main statistically significant climate-driver of streamflow change. The contributions of other non-climate factors, such as the change in land cover, to the reduction in annual streamflows are shown to be minor in comparison.

**Keywords**

Long-term statistical trends, basin-scale hydrology, modelling, seasonal distribution

# 1. Introduction

According to the Intergovernmental Panel on Climate Change (IPCC), the Mediterranean Basin is a particularly sensitive region to changes in the climate, where the effects of global warming are already evident and more pronounced than in the rest of Europe (IPCC, 2007). Precipitation in the last fifty years, for example, has decreased significantly, and by the end of the XXI century it will undergo a 20% decline, and up to 45% in summer. The temperature has also increased and will continue to grow faster than the European average (IPCC, 2007). As a result of these changes in temperature and precipitation, water resources available for agriculture, industry and urban water supply in the Mediterranean basin, already scarce and mainly dependent on streamflow (Viviroli and Weingartner, 2004; De Jong *et al*., 2009), are expected to decrease (García-Ruiz, *et al*., 2011). Global scale simulations of river flows conducted with several climate models (Milly, *et al*., 2005), for example, predict a 10–30% decreases in streamflow in southern Europe and the Middle East by the year 2050. River streamflows are expected to decrease even more dramatically in summer time, with reductions of up to 80% by the end of the century (IPCC, 2007). Thus, intra-annual flow patterns are also expected to undergo significant changes (García-Ruiz, *et al*., 2011; Blöschl et al., 2017). Being able to detect these trends at the river basin-scale is of paramount importance for water managers, who should be able to recognize if their data, upon which the design and operation of water resource systems were based, are no longer consistent with current conditions (Lins and Cohn, 2011). This is a particularly critical task in semi-arid Mediterranean river-basins, where a subtle and difficult-to-manage equilibrium between water availability and demand exists. In the Guadalquivir River Basin in Southern Spain, for example, the average water resource availability is 7,043 Mm$^3$/yr, but with a range from 372 to 21,530 Mm$^3$/yr. Total water demands, in turn, are stable and around 3,568 Mm$^3$/yr. Hence, and as a result of the hydrologic variability, there will be years with severe water deficits (Berbel *et al*., 2012). Any long-term changes in annual streamflow volumes in basins, like the Guadalquivir Basin, with severe and structural "water deficits" (Bhat and Blomquist, 2004) will likely aggravate the challenge of allocating scarce water supplies. Hence, there is an urgent need to detect and understand changes in streamflow volumes or flow regimes, so that they can be accounted for in water resources planning and management. The analysis and detection of trends in the streamflows of large river basins, however, should be carried out with care, given that those trends may arise as a result of land-use changes, changes in water infrastructure and other factors (García-Ruiz, *et al*., 2011, Asarian and Walker, 2016) of anthropic origin, but not directly linked to climate change. This makes the analysis of hydrologic trends in large basins a challenging task.

A large number of studies have been conducted worldwide to assess the impact of climate change on hydrological variables during the most recent decades. These include the studies of Mohsin and Gough (2010) in North America, Yang and Tian (2009) in Asia, Nasri and Modarres (2009) in the Middle-East, or that of Del Río *et al*. (2011) in Southern Europe, on the analysis of trends in precipitation and temperatures; or the work of Zhang *et al.* (2011) in China, or Biggs and Atkinson (2011) in the British Islands on trends in streamflow volumes, just to mention some recent examples. Few of these studies, however, have analyzed trends in river streamflow volumes and flow regimes driven by climate change in river basins of the Iberian Peninsula. The work of Morán-Tejeda *et al*. (2011) is one of those few studies conducted at the river basin scale, in the Duero River Basin. They analyzed, one by one, monthly records collected during the last four decades in a total of 56 gauging stations throughout the basin, and found a marked reduction in river discharge in winter and spring. They attributed those changes in discharge to decreasing winter precipitation and a reduction in the snowmelt component, presumably due to increasing temperatures in winter and spring (Moran-Tejeda, *et al*., 2011). Estrela *et al*. (2012) reviewed recent modeling studies and expected impacts of climate change on water resources in Spain. According to their study, streamflow reductions are expected to be between 10 and 30% through the XXIst century, being the most critical areas the southern and southeastern arid and semi-arid regions, where water scarcity and drought problems are particularly acute and recurrent. Lorenzo-Lacruz, *et al.* (2012) analyzed the trends in streamflow volumes at the regional scale of the Iberian Peninsula, and also found a generalized and significant decreasing trend in streamflow during winter and spring. The magnitude of those trends varied depending on the basin, but were more pronounced in the central and southern basins (Tajo, Júcar, Guadiana and Guadalquivir). The trends in streamflows were partly attributed in that work to decreasing trends in precipitation, but also to reforestation and increasing water demands (López-Moreno *et al*., 2011). From the work of Lorenzo-Lacruz, *et al*. (2012) it is not clear, though, whether the changes in monthly streamflow volumes during the last decades are driven or not by changes in hydrologic forcing, or what is the contribution of the long-term global scale variability in climatic forcing compared with local-scale changes in water demands or land-use, as drivers of the changes in streamflow. Being able to differentiate between climate and non-climate drivers of streamflow changes is important from a water management point of view, since the former, being global processes, are not controllable at the basin scale, hence, they cannot be managed. Thus, a deeper analysis should be conducted to obtain a confident picture of the water resources available (Lorenzo-Lacruz, *et al*., 2012), and to develop a sound understanding of the factors (climatic and non-climatic) driving long-term changes in streamflow volumes and flow regimes.

Our goal is to detect and understand climate-driven basin-scale trends in streamflows over the recent decades in the Guadalquivir River Basin, the southernmost and the most arid of the five large basins of the Iberian Peninsula (Fig. 1a). Our work is based first on the statistical analysis of monthly streamflow records from a selected network of gauging stations in the Guadalquivir River Basin, and a gridded set of monthly precipitation, $P$, and potential evapotranspiration, $ET_0$, data covering the southern end of Spain, to detect the existence of and quantify climate-driven trends in the basin-scale hydrological response during the last half of the XXth century (from 1950 to 2005, study period). The large-scale coherent and synchronized signals underlying the hydrologic information in the database, representing the regional hydrologic behaviour of the basin, was isolated using a Principal Component Analysis, and their trends evaluated and interpreted. A lumped-parameter physically based continuous model was used to evaluate the relative contribution of climate and non-climate factors to the observed changes in streamflow. The flow records used in this work were carefully selected following the same criteria used in the literature to identify reference hydrologic networks (see Whitfield *et al*. 2012 and references there in), which include (1) the existence of long and trustful records, (2) the absence of upstream hydraulic structures, and (2) weak or negligible land-use changes.

Our work differs from earlier publications dealing with long-term streamflow trends in the Iberian Peninsula in several aspects. First, from a methodological point of view, since we isolate and examine regional time-coherent streamflow trends, besides conducting the analysis on a station-by-station basis. The time-coherent regional signal is constructed using streamflow records from a reference hydrologic network, consisting of a series of stations in which we ensured that the human impact was minimal. This is in contrast to previous work that included gauging information from both regulated and un-regulated rivers. With this approach, we are aiming at isolating the driving effect of a changing climate on basin-scale streamflows, which, by definition, should have a large fingerprint, coherent in time, by minimizing the possible effects of changes in land-use or water withdrawals that might contaminate the statistical analysis and makes the identification and interpretation of climate-driven signals difficult. Any signal constructed from the reference network of stations will be assumed to be climate-driven. Hence, and from here on, the term "streamflows" will refer to climate-driven streamflows. The trends in our streamflow signals will likely not represent the behaviour of other gages where the anthropic influence is strong. Using a reference hydrologic network to detect climate-driven trends in streamflows introduces some uncertainties in the analysis. A well-known pitfall of this approach has to do with the non-uniform spatial distribution of the nearly-pristine gauging stations, which tend to be concentrated in the headlands of the river basin. However, we will demonstrate that the regional signals are consistent, independent of the number of stations used to reconstruct it varying from 8 to 43. Moreover, and to make sure that the non-uniform and sparse distribution of stations does not introduce any bias in our conclusions, a gridded-data set of streamflows covering the whole basin and the period from 1950 to 2005, was constructed with the model from the gridded P- and $ET_0$ data set, and analyzed. This gridded set was assumed to represent the basin-wide streamflow signal, and, any of the trends identified in it can only be atttributed to changes in hydrologic-forcing, and not to other non-climatic anthropic factors. But, what makes our study novel compared with prior work in the Iberian Peninsula is the attribution work which is conducted using both data-based and simulation-based methods (see Merz *et al.* 2012).

Not only we describe and detect trends in streamflows, but we assess the contribution of different changes in hydro-meteorological forcing detected in the data set. The attribution work is also used to verify the assumed unimportance of non-climatic factors as drivers of changes in the streamflow data set. Hence, ours is a sound and careful analysis exercise, that focuses on a particular basin in southern Spain, of large dimensions, which has not been studied in this detail before, but has distinct hydrological regimes and the atmospheric/climate forcing when compared to other basins in the Iberian Peninsula, as shown for example in the work of Hidalgo-Muñoz *et al*. (2015).

## 2. Methods

***Study site***.- With a surface area of 57185 km$^2$, the Guadalquivir Basin (Fig. 1a) is characterized by exhibiting large temporal and spatial variations in hydrometeorological conditions. Average annual *ET*$_0$, for example, decrease from the nearly 1400 mm/year, in the low altitudes of the valley on the W- to ca. 900 mm on the SE- and E- of the basin. Annual precipitation values also exhibit large spatial variations with minimum values of ca. 300 mm in the SE- and N of the basin and maximum values which are almost twice the minimum values in NE and NW. Annual precipitation averaged over the basin ranges from as little as 300 mm to as much as 1100 mm, depending on the year (Peña-Gallardo *et al*., 2016). Furthermore, years of extremely high or low precipitation tend to cluster together, compounding the effects of droughts or flood. Most precipitation is concentrated in the winter months, with peak rainfall occurring from November through March. Precipitation is virtually nil during the long and dry summers, when potential evapotranspiration peaks (Argüeso *et al*., 2011, 2012).

Permeable limestones and dolomites are predominant in the left margin of the basin, particularly in the E and SE. The right margin, in turn, is characterized by the existence of largely impermeable lithologies (see Fig. 1b).

*Hydrologic data set* - Monthly streamflow data from more than 230 gauging stations in the Guadalquivir basin were provided by the Spanish Center for Public Work Experimentation and Study (CEDEX, *Centro de Estudios y Experimentación de Obras Públicas*). Out of the 230 gauging stations in the original data set, we selected a total of 57 that were not affected by upstream hydraulic infrastructures. The homogeneity of the streamflow records in this reduced data set of 57 gauging stations was checked as in Hidalgo-Muñoz et al. (2015). Data gaps were filled using linear regression models from records collected at neighbouring stations having correlation coefficients R $\geq$ 0.7. In only eight of these stations (marked in Fig. 1a) the percentage of filled data in their 56-year time series was below 20%; in only six if them the percentage of gaps was below 5%. We will use the records from these eight stations in the analysis of long-term trends (**long-term reference stations**). The number of stations with a percentage of gaps in their records below 5% and 20% varied depending on the sub-period of time considered (see Table 1). Note that only 43, out of the 57 gauging stations not affected by water infrastructures, have records in the period from 1995 to 2005, with at most 20% of gaps. These 43 watersheds will be used as the **full reference data set** when analyzing the spatial consistency of the streamflow signals (shown in Fig. 2). Table 2 gathers the following information for each of these watersheds: code and name of the station, surface ($km^2$), average elevation (m) and the mean annual precipitation (mm) and potential evapotranspiration (mm) during the study period. For all the calculations, we consider hydrologic years, starting in October and ending in September. Fall is considered to start in October and finish in December; Winter will include the three months from January to March; Spring, from April to June; and, Summer from July to September.

Monthly rainfall $P$ and temperature $T$ data for the Guadalquivir River Basin during the study period were extracted from a long-term high-resolution (~ 20 x 20 km) gridded dataset SPAIN02 covering Spain (Herrera *et al.*, 2012), and provided by the Spanish Meteorological Agency (AEMet). The grid is the result of interpolating in space daily precipitation records from more than 2000 stations and daily averaged maximum and minimum temperature records from more than 250 stations in Spain and Balearic Islands. The gridding methodology used for precipitation is based on a two step krigging approach. In the case of temperatures, thin plane splines are fitted to the monthly data considering elevation and an ordinary krigging was later applied to the residuals (Herrera *et al.,* 2012). When comparing different simplified methods to calculate evapotranspiration, $ET_0$, in the Iberian Peninsula, Vicente-Serrano et al. (2014) found that the Hargreaves equation (e.g. Hargreaves and Samani, 1982) was the most adequate of the methods analyzed, representing the same tendencies as encountered with more sophisticated approaches such as the Penman-Monteith's equation. However, Vanderlinden et al. (2004, 2008) showed that the Hargreaves method produces noticeable bias in Southern Spain, and proposed a modification of the original equation through the adjusted Hargreaves coefficient (AHC). In this work the Hargreaves equation together with the AHC were used to estimate $ET_0$ from the maximum and minimum temperature data at each grid point.

*Statistical analysis*.- Principal Component Analysis PCA (Jollife, 1986; Jackson, 1991) was applied separately to each variable in the hydrological data set to isolate the large basin-scale or regional signal. Prior to the analysis and given the strong seasonal variability of the Mediterranean climate, the annual cycle was removed from the hydrological data by subtracting the averaged monthly values from the raw records. The large-scale regional signal was taken as the amplitude of the most significant principal components, representing the coherent and synchronized variations of hydrological information in all the stations included in the PC analysis. In studying the streamflow data, the PC analysis was applied to different subsets of the data, all of them ending in 2005, but ranging in length from 16 to 56 years (1950-2005, 1960-2005, 1970-2005, 1980-2005 and 1990-2005). A varying number of gauging stations were included in those sub-sets, all of them with at most 20% of their data filled (see Table 1). Our goal, by doing these exercises, was to demonstrate that our conclusions, regarding streamflow trends, are robust and independent of the length of time and number of stations included in the analysis.

As in Salmi *et al.* 2002 (see also Drápela *et al*., 2011; or Mondal *et al*. 2012), the Mann-Kendall (MK) non-parametric statistical test (Sneyers, 1975) was used here to establish whether the time series of the basin-scale hydrological signals correspond to a stable climate, characterized by a simple random series (Null hypothesis, $H_0$), or, if in turn, the series has statistically significant trends. Quantitative estimates of the temporal trends in hydrological variables were, in turn, obtained using non-parametric Sen's slope methods (Sen, 1968). The trend analysis (MK test and Sen's slopes) was first conducted with the regionalized annual precipitation, $ET_0$ and streamflow data set. It was, then, applied on a season-by-season basis, to check seasonal trends. The sequential version of the MK test (SMK, Sneyers, 1975; Esteban-Parra *et al*., 1995) was also applied to these seasonal series. The impact of serial correlation on the results of the Mann-Kendall test has been addressed as suggested by Kulkarni and von Storch (1995). In addition, the trend analysis was also conducted on a site-by-site basis, studying records from individual stations, one by one.

*Changes in land-cover* - Land-use changes in the basin were characterized through a series of maps representing the prevailing land-use in 1956, 1977, 1984 and 1999, available through the Environmental Information Service of the Regional Government of Southern Spain (http://www.juntadeandalucia.es/medioambiente/site/rediam/informacionambiental). Similar categories to those employed in the CORINE Land Cover project are adopted to classify land-use in these maps. From this series of land-cover maps, and the 20 x 20 m resolution Digital Elevation Model of the Spanish National Geographic Institute (http://www.ign.es/ign/main/index.do), we constructed four raster 20 x 20 m resolution maps of Hydrologic Curve Numbers (CN) for average soil moisture conditions, following the same approach as proposed by Ferrer (2003). The map of soil hydrological groups used to construct the CN maps was provided by Ferrer (personal communication). The Hydrological Curve Number CN is a parameter commonly used in the literature to estimate direct runoff and groundwater infiltration from rainfall excess (e.g. Ponce and Hawkings 1996), and the changes experienced by CN in the period from 1956 to 1999 were used to quantify the hydrological effect of land-use changes. An average CN for each of the 43 reference watersheds and for each year (1956, 1977, 1984 and 1999) were then calculated by, first delimiting those watersheds, and averaging the CN data within those limits. For presentation purposes, to describe the changes occurring in the landscape from 1956 to 1999 we used a series of reclassified land-cover maps, in which the original categories from the CORINE Land Cover project were reclassified into four land uses. These land uses include 'Cultivated Lands', 'Grasslands' and 'Woods and Forests' (the three main categories identified in the National Engineering Handbook, USDA, 2002), and a fourth category referred to as "Others" and corresponding to pervious rocks and impervious surfaces.

*Hydrologic modelling* - A model was calibrated to simulate the hydrologic response of two specific watersheds in the Guadalquivir River Basin during the study period, and, then, used to assess the relative impact of changes of varying intensity in climatic and non-climatic factors on streamflow. The watersheds simulated were those contributing to two reservoirs: Tranco de Beas (6001) and Cala (6014). They were chosen for this exercise for several reasons: (1) its surface area is large, and satisfies the area criteria for the application of the model proposed by Témez (1977) and other users; and (2) the data set for these watersheds had less than 5% of gaps in the period 1950-2005. Furthermore, in the statistical analysis, it was shown that the discharge records from these two stations represented well the regional basin-scale streamflow signals, given their high-correlation with the first principal component, but also exhibited the most contrasting sub-regional large scale hydrological behaviors, with the maximum positive and negative correlation with the second principal component.

The lumped-parameter physically based continuous model proposed by Témez (1977) and extensively used in hydrological studies of the Iberian Peninsula (Cancela *et al.*, 2004, Bejarano *et al.*, 2010) was adopted for this modeling exercise. The simulations are driven by monthly-series of precipitation $P$ and potential evapotranspiration $ET_0$. Water in the basin is stored in two different buckets or reservoirs, representing groundwater ($G$) and soil moisture ($H$). The latter has a limited capacity, $H_{max}$. As long as the soil is not saturated at the beginning of a time step (i.e. $H < H_{max}$), a fraction of the rainfall will be stored as soil moisture. The remaining is referred to as surplus ($S$), which either becomes surface streamflow ($F$) or is stored as groundwater ($G$). The surplus only occurs as long as precipitation exceeds a threshold $P_0$, which is proportional to the moisture deficit ($H_{max}$ - $H$). Streamflow leaves the basin on the same month that is generated. The fraction of the surplus stored as groundwater becomes the base-flow component of the river discharge (base-flow) at later times. The equations used to update the state variables at any given time $i$, given their values at the previous time step $i$-1, are shown in Appendix I.

A total of four parameters are used to account for the characteristics of the watershed in the model. These parameters include (1) the maximum soil moisture content, $H_{max}$; (2) the maximum groundwater volume stored in the aquifer, $I_{max}$; (3) a parameter, $\beta$, that quantifies the rate at which groundwater feeds streamflow; and, (4) a reducer coefficient, $C$, depending on the soil type, vegetation and landscape morphology, that is used to calculate the threshold precipitation from the moisture deficit. Being the conceptualization of the watershed similar to that used in the event-based Curve Number model developed by the U.S. Soil Conservation Service (SCS) one can establish parallelisms between the model parameters of these two models (see Appendix I). In particular, $H_{max}$, can be considered equal to the maximum storage, $S$, in the SCS model (e.g. Ponce and Hawkings 1996), and hence, it can be shown to be inversely proportional to the Hydrologic Curve Number, CN. Curve numbers, in turn, were calculated for different years from the series of land cover maps as indicated above. Of the remaining parameters, $C$ was set to 0.2 as in the SCS model, and the other two were calibrated using the Shuffled-Complex-Evolution Algorithm of Duan *et al.* (1994). The Mean Squared Error (MSE) was chosen as the objective function in the calibration process. Once the objective function was minimized, the goodness of the fit was evaluated by carrying out a linear regression analysis between the observed and the simulated streamflows, determining the value of the coefficient of determination $R^2$ for the two selected watersheds. The allowable range of values for those adjustable parameters in the calibration exercise were those recommended in Martos-Rosillo *et al.* (2006) and Murillo and Navarro (2011) for southern Spain. The monthly rainfall and potential evapotranspiration data used to force the model was interpolated from the gridded $P$ and $ET_0$ data set (see section *Hydrologic data set*). Thiessen polygons were used for data interpolation.

*Sensitivity analysis* - A sensitivity analysis exercise was conducted with the calibrated model to assess the relative effect of observed changes in hydrologic forcing and basin characteristics on streamflow $Q$ at regional scale. The sensitivity of the model streamflow was assessed in this exercise using a First-Order Variance Analysis FOVA (e.g. Blumberg and Georgas, 2008). The reference simulation was conducted using the calibrated values of the model parameters for each watershed, and was forced using a 56-year long stationary series of precipitation, $P_{ref}$, and evapo-transpiration, $ET_0^{ref}$, constructed as follows. First, a monthly time series of $P$ and $ET_0$ for an average year was constructed from the observed series. The observed series will be referred to as $X_{ij}$ where, $X$ is either $P$ or $ET_0$, the index $j$ ($j = 1, …, N$) identifies the year in the time series, and, $i$ ($i = 1,…, 12$) the month of the year. The monthly records for the average year, $\underline{X}_i$, was calculated as:

$$\underline{X}_i = \frac{1}{N}\sum_{j=1}^{N} X_{ij} \qquad (i = 1,…, 12) \tag{1}$$

The monthly series $P_{ref}$ and $ET_0^{ref}$ were constructed by repeating $N$ times the intra-annual pattern represented by Eq. 1. Note that the series $P_{ref}$ and $ET_0^{ref}$ used to force the reference simulations will vary depending on the particular watershed. The model was then run with perturbed hydrologic conditions in the watershed, and the sensitivity of the model streamflow to those perturbations was quantified with a dimensionless sensitivity coefficient $S_p$,

$$S_p = \frac{\Delta Q / Q_0}{\Delta p / p_0} \tag{2}$$

In this expression, $Q_0$ is the annual streamflow in the reference simulation, $\Delta p$ is a sufficiently small change or perturbation in either $H_{max}$ or the forcing series relative to the reference or baseline value $p_0$, and $\Delta Q$ is the change in annual streamflow that results from those perturbations. The coefficient of variation of the streamflow $CV_Q$ measuring the relative change in annual streamflow in response to any given changes in the hydrologic conditions of the watershed can be calculated as follows (e.g. Blumberg and Georgas 2008):

$$CV_Q^2 = \sum_{p=1}^{n} S_p^2 CV_p^2 \tag{3}$$

Here $CV_p$ is the coefficient of variation of either $H_{max}$ or hydrologic forcing, and $n$ is the number of parameters used to represent the changes in hydrologic conditions experienced in the watersheds, and identified in the statistical analysis of the full and long-term reference data sets. Note that the contributions from different parameters to changes in streamflow are additive in this first order approach. Once all the values of $S_p$ were calculated, $CV_p$ was determined for $H_{max}$ and the forcing series. In the case of $H_{max}$, the coefficient of variation is calculated as follows:

$$CV_{Hmax} = \frac{\Delta H_{max}}{H_{max-0}} = \frac{H_{max}^{1999} - H_{max}^{1956}}{H_{max}^{1956}} \qquad (4)$$

Where $H_{max}^{1956}$ and $H_{max}^{1999}$ are the values of $H_{max}$ in a given basin for the years 1956 and 1999, respectively. Therefore, $CV_{Hmax}$ expresses the relative change in the parameter $H_{max}$ in the study period. As analyzing $CV_p$ for the forcing series, it is calculated from the observed trends in the first principal component of the hydrologic variable studied in each case:

$$CV_p = \frac{\Delta p}{p_0} = Sen(PC1(p)) \cdot L_p \qquad (5)$$

In this equation, $Sen(PC1(p))$ is the Sen's slope of the first principal component of the hydrologic variable represented by the forcing series and $L_p$ is the length of the forcing series. In this case, $CV_p$ expresses the percentage of change of a given forcing series in the study period. The different values of $S_p$ and $CV_p$ allowed us to study the change in streamflow associated $H_{max}$ and the forcing series individually and the change in streamflow as the result of the changes in all of them simultaneously using Eq. 3.

**3. Results and Discussion**

***Regionalized hydrological records -*** Two significant Principal Components are encountered in the analysis of the precipitation data. Their loading factors and time series are shown in Figs. 2a-b and Fig. 2f-g, respectively. The first principal component PC1(*P*) explains 72.6 % of the total variance and represents the seasonal pattern characterized by rainy winters (Fig. 2k). Having high positive and significant correlation (or loadings) in all the grid points (Fig. 2a), this component, represents the regional scale changes in precipitation. The second PC, PC2(*P*), with an explained variance of 6.7 %, presents only significantly positive loadings over the SE of the basin, where precipitation tends to remain high from fall to spring time (Fig. 2l). Hence, it represents sub-regional large scale variations in the hydrological forcing. Only one significant Principal Component, PC1($ET_0$) was detected in the analysis of the $ET_0$ data set, using the Scree Plot test, that explains around 60% of the total variance. Figure 2c shows the loadings for this first PC, with very high correlations in all the grid points. This PC1($ET_0$) is linked with the thermal regime of the region, characterized by minimum temperatures in winter, growing up to reach the maximum values in summer time (Fig. 2m).

The PC analysis was applied to different subsets of streamflow data, varying in length from 16 to 56 years (1950-2005, 1960-2005, 1970-2005, 1980-2005 and 1990-2005) and varying, also, in the number of stations included (8, 10, 16, 16 and 34, respectively), but in all cases, with at most 20% of their data filled. Independently of the subset used, the first two principal components of the streamflow data, PC1(*Q*) and PC2(*Q*), were the only significant PCs and explained more than 70% of their variance, even, reaching 84% for the period 1990-2005. Hence, we will only use these two significant PCs to characterize streamflow at the regional (basin) scale. The loading factors for those first two PCs that arise in the analysis of the full reference data set are shown in Figs. 2d-e.

The first principal component PC1($Q$) explains more than 73 % of the variance in the data sets analyzed presenting high loadings in all stations, most of them close 1 (Fig. 2d). It describes the most common streamflow pattern of the Guadalquivir Basin, with streamflow increasing from October to January-February and decreasing thereafter to reach their minimum values in summer time, from July to September (Fig. 2n). Hence, it represents the regional (or global) response to rainfall. The second principal component PC2($Q$) explains 9 % of the variance in the streamflow data and has a more complex spatial correlation than PC1($Q$). Maximum positive loadings ($R \sim + 0.5$) occur in the SE of the basin and maximum negative loadings ($R \sim - 0.4$) in W- and NW- (Fig. 2e). PC2($Q$) describes the large scale sub-regional response to precipitation, with maximum streamflows occurring early in spring time (April–May) and lower values in late fall (November-December) (Fig. 2o). This is the seasonal pattern characterizing streamflows in snow-melt driven watersheds. But, this pattern can be also interpreted as the result of the delayed response to precipitation of the groundwater fed base-flow component of the discharge. Morán-Tejeda *et al*. (2011) opted for attributing the peaks in spring streamflow to snowmelt, and justified their attribution based on the high elevations of the contributing watersheds in the Duero Basin. In our case, though, only five of the watersheds included in the full reference set, draining the highest elevations of Sierra Nevada (see Fig. 1a), exhibit temperatures in winter that are persistently below $0^o$C. Only in these watersheds the contribution of snowmelt processes to streamflow is significant. In all others watersheds in the E and SE of the basin, with a significantly positive correlation with PC2($Q$), average temperatures in winter are above $0^o$C (see the maps of Ninyerola et al. 2005 in http://www.opengis. uab.es/wms/iberia/index.htm).

The attribution of the spring peaks in streamflow in these watersheds can only be attributed to the delayed response of groundwater flow to precipitation. This interpretation is supported by the fact that the areas of higher permeability are located in left margin of the basin (Fig. 1b). Moreover, precipitation in this area tends to remain higher, when compared to the NW, until the month of April, as revealed by the positive correlation exhibited by PC2($P$) with the grid points sited in the E-SE of the basin (see Fig. 2b and 2l).

The mean annual PCs series of streamflow for the different subsets of data, varying in the number of stations used, are shown in Fig. 3. Note that for the common periods in the datasets, both PC1($Q$) and PC2($Q$) are very similar, which suggests that the regional or basin-scale signal is very robust and can be reconstructed from the analysis of a limited number of stations, even if they are not uniformly distributed. So, the PCs series for the period 1950-2005 calculated using the eight stations in the long-term reference set, with the largest unfilled records (see Fig. 1a), can be considered, at least for the purpose of analyzing trends, representative of the long-term, time-coherent, basin scale climate-driven streamflows. The loading factors for those reference stations associated to the first two PCs are shown in Table 3. Note that the two stations with largest negative and positive loading with PC2(Q) are 6001 and 6014, respectively. Hence, these two stations can be considered as representing the largest contrast in the large scale sub-regional hydrologic behavior.

***Long-term trends in regionalized hydrological variables .-*** Regional-scale streamflows, as characterized by PC1($Q$), exhibited significant decreasing trends, both when analyzing the full data set, considering all months, or, when applying it on a season-by-season basis (Table 4). PC2($Q$) also exhibited significantly negative trends.

Given the positive signs of the loading factors in the SE of the basin (negative in the NW, Fig. 2e), the negative trend in PC2($Q$) implies that the decreasing trends in streamflows are likely stronger in the SE compared to the rest of the basin. These significant and negative trends in PC2($Q$) also occur when analyzing the data set on a season-by-season basis (Table 4). Lorenzo-Lacruz *et al.* (2012), working at the scale of the Iberian Peninsula, also found that the streamflows exhibited significant decreasing trends in winter and spring (see also the work of Morán-Tejada *et al.*, 2011, in the Duero River Basin), but they reported increasing streamflow trends in summer and fall. But note that both Morán-Tejada *et al.* (2011) and Lorenzo-Lacruz *et al.* (2012) included in their data sets gauging stations sited downstream of major regulation infrastructures. In fact, Lorenzo-Lacruz *et al.* (2012) attributed the positive trends in summer and fall to water management strategies. In any case, our results are consistent with previous reports, in that streamflows, in general, have decreased in the last few decades (Stahl *et al.* 2010, Milly *et al.*, 2005, or López-Moreno *et al.* 2011).

Regional-scale precipitation, as revealed by PC1($P$) (Fig. 2f), exhibited a negative but not significant trend during the study period. The trends of the sub-regional signal, PC2($P$), was also negative but, in this case, it was significant. Given the different signs of the loading factors (positive in the E-SE and negative in the NW Fig. 2b), this implies decreasing precipitation levels over the E-SE, since 1950, and increasing values over the NW. Compared to the annual data, the changes in the regional-scale precipitation were stronger when considered on a season-by-season basis, but only in winter they were statistically significant. Trends in winter precipitation were particularly strong and negative, compared to the other seasons, when precipitation tends to exhibit weaker, and even positive, trends (Table 4). These decreasing trends in winter rainfall, as previously reported (Xoplaki *et al.,* 2004; Mourato *et al.* 2009) can be used to explain the decrease in winter streamflow. The decreasing trends in the streamflows in other seasons, however, cannot be explained in terms of rainfall. Instead, they can be attributed, in part, to increases in evapotranspiration losses. Regional scale evapotranspiration, as represented by PC1($ET_0$) in Fig. 2h, exhibited a positive trend, significant at 95% confidence level. These trends were also significant in winter and spring (Table 4), with changes of up to 25%. These results are consistent with those of Brunet *et al.* (2007), for the Iberian Peninsula.

*Spatial distribution of long-term trends: station-by-station analysis* - Sen's slopes for seasonal and annual precipitation at each grid point in the basin are shown in Fig. 4f-j. Annual precipitation does not present significant trends, except over the E-SE regions, where significant decreases of ca. 8 mm year$^{-1}$ (2%year$^{-1}$) are detected. This decrease is mainly associated with the significant and negative trends detected in winter over most of the basin. In the E- end of the basin, negative and significant trends are also detected in fall. The positive trends detected over the W- in summer time is not important given that precipitation in this season is negligible. The trends in seasonal and annual $ET_0$ records are shown in Fig. 4a-e. Those trends are significant and positive trends over the NW and SE of the basin, and can be of up 6 mm year$^{-1}$ (< 1%year$^{-1}$) in some locations (Fig. 4e). Winter $ET_0$ is characterized by a general increase, significant at 95% confidence level, over most of the basin. Significant trends are also found for the other seasons over the NW and SE of the basin. It is remarkable that the magnitude of these trends is lower than those encountered in precipitation. For example, in the SE, where the trends appear to be stronger, annual precipitation exhibits trends of ca. -1.5%year$^{-1}$, $ET_0$ only shows increases of 0.2%year$^{-1}$.

Sen's slopes (expressed as percentage respect to the median) of the annual streamflow recorded at the long-term reference stations from 1950 to 2005 are shown in Table 5. Sen's slopes of $P$ and $ET_0$ records interpolated to the contributing watersheds of those gauging stations using Thiessen polygons are also shown in Table 5. Note that streamflows exhibit negative trends in all the stations. But those trends are significant only in the E-SE portion of the basin. The streamflows recorded in the E- exhibit trends of 1 to 2 %·year$^{-1}$, and have a level of significance of at least 90%. It is remarkable that these negative trends are found in watersheds where the annual precipitation does not present significant changes. That is the case of Rumblar, Guadalmellato or El Doctor. In Tranco de Beas and Cubillas, both annual streamflows and precipitation exhibit decreasing trends, with at least 90% of significance.

Seasonal streamflow and precipitation (Table 5) also exhibit decreasing trends in most cases, being predominantly significant at 90% level for winter and spring. $ET_0$ also increases in most cases, but those changes are particularly significant and generalized in winter. Trends in other seasons and even at annual scales are not generalized. Only in three of the long-term reference stations, annual $ET_0$ exhibited significant trends, in all cases, positive. Two of them (stations 6011 and 6014) are sited in the W of the basin. In these two stations, seasonal $ET_0$ increases significantly in summer and spring. The decreasing trends in annual streamflow are even more evident when analyzing the Sen's slope for the period 1960-2005 (see Fig. 3a). In this case, the trends can be of up to 4%·year$^{-1}$. For this period, the negative trends for the winter streamflow present values close 5%·year$^{-1}$.

These results suggest that the records can be grouped in two classes, depending on whether they correspond to gauging stations located in the NW or in the E-SE. Stations in the E-SE exhibit particularly strong and significant trends in annual streamflow, compared to those exhibited in the NW. Those changes in the E are likely linked to changes in annual precipitation. But the largest and most significant changes detected in this area tend to occur in winter and spring streamflow, winter precipitation, and winter $ET_0$. The differential behavior exhibited by E-SE and NW stations is captured by the sub-regional PC2 for streamflow and precipitation.

*Climate factors contributing to decreasing streamflow* - Our analysis of the long-term regionalized hydrological variables suggests that streamflows during the last decades in the Guadalquivir River basin have exhibited a statistically significant negative trend, in all seasons. Here, we propose a plausible explanation to the decreasing trends in streamflow based on the analysis of rainfall and $ET_0$ data considered on a season-by-season basis. The time series of PC1($P$), PC1($ET_0$) and PC1($Q$) for each season are represented in Fig. 5. The seasonal trends can be detected in those plots, but they are more clearly appreciated in the analysis of the results of the SMK test, shown in Fig. 6. From 1960's to 1980's, streamflows in fall have experienced a marked decreasing trend, likely as a result of (1) decreasing values of precipitation and (2) increasing evapotranspiration losses (Fig. 6a). During the last 20 years (from 1980's to 2000's), in turn, fall streamflow, precipitation and $ET_0$ have remained nearly constant with no significant trends. The changes in fall streamflow appear to be largely explained in terms of fall precipitation with which the correlation is large and significant (R ≈ 0.92). The correlation of fall streamflow with fall $ET_0$ is lower and not significant, though (R ≈ -0.30). Furthermore, the correlations of fall streamflows with the hydroclimatological conditions prevailing in summer are also not significant.

Streamflow values in winter have exhibited, in general, a strong decreasing trend during the study period (Fig. 6b). These changes are largely driven by the reduction in winter rainfall (Fig. 6b). There is no significant correlation between winter streamflow and $ET_0$ (R≈ -0.40), but, the linear correlation between winter precipitation and streamflow is high and significant (R ≈ 0.84, see Fig. 7a). Note that some points in Fig. 7a are far from the regression line. Some of those outliers are indicative of very rainy winters and low streamflows. Those winters are typically preceded by very dry falls. Soil does not become saturated at the start of the winter, on those years, and the first winter rains are used to increase the soil moisture content to saturation, rather than to generate streamflow. Hence, winter streamflows are significantly affected by fall precipitation. In fact, the correlation between fall precipitation and residual winter streamflow is significant R≈ 0.67 (Figure 7b). The tendency of winter $ET_0$ to increase, as revealed in Fig. 6, is low compared with the observed reduction in rainfall and appears to have a minor effect on the streamflows, compared to the effects of decreasing precipitation (see Table 3 and Fig. 4).

A marked reduction in spring runoff have been recorded in the Guadalquivir River Basin during the last three decades (Figs. 5c and 6c). Given that precipitation has not changed significantly in spring time (Fig. 6c), this reduction is largely driven by increases in $ET_0$ (see also Table 4). Note also the decreasing trend in spring streamflow starts in the late 70s, coinciding with the onset of the increasing trend in $ET_0$. The linear correlation between spring streamflow and $ET_0$ is significant, in contrast with other seasons (R ≈ -0.53), but still is low, which suggests that other factors probably need to be taken into account to understand the decrease in spring streamflow. In particular, one needs to consider the prevailing hydrologic conditions in winter which may affect the streamflows in spring time. The relationship between winter and spring conditions can be better understood if we plot and compare the time series of winter and spring precipitation, and spring streamflows (Fig. 8). Note that the high values of streamflow recorded in spring time are not necessarily associated to large values of spring rainfall. For example, in 1954, 1959 or 1979, among others, rainfall in spring is low compared with the streamflows. In those years, the high spring streamflow values are, in turn, associated to high winter precipitation. Inversely, the low spring streamflow values of 1960 and 1999 can be associated with previous dry winter conditions. The influence of winter precipitation on spring runoff can be partly attributed to the dynamics of snow accumulation and melting, but, in the Guadalquivir Basin, it is more likely the result of the slower response to rainfall of the groundwater fed base-flow component of discharge.

Rainfall during summer time has tended to increase in the last four decades in the Guadalquivir Basin (Fig. 6), but this increase has not lead to increases in streamflows. Summer time $ET_0$ (ca. 180 mm) is almost six times the average precipitation (29 mm). Hence, any additional rainfall in summer will be evaporated, being the excess rainfall zero at all times. Streamflow records in summer time largely represent the base-flow component of runoff, with an important groundwater contribution. Hence, they are largely determined by the hydrological history of the basin during the previous seasons.

*Land-use changes in the basin* - The percentages of the area of each basin occupied by each of the four land-use categories considered in the reclassified maps are shown in Table 6, both for 1956 and 1999. Note that 'woods and forest' represent the most common land-use in most watersheds, representing, on average, almost 60% of their surface. 'Cultivated lands', including arable crops such as wheat or rye, olive groves, vineyards and almond trees, on the other hand, represent 25% of the area. Of the agricultural lands, only 2% is irrigated, the remaining being rainfed. From 1956 to 1999, irrigated lands have increased, occupying in 1999 almost 1% of the area of the reference watersheds. The cultivated lands, including both irrigated and rainfed, though, have decreased in the 40 years from 1956 to 1999. In contrast, woodland and forest have increased in almost 50% of the watersheds analyzed. Previous studies conducted in other basins in the Iberian Peninsula suggested that the decrease of streamflow detected in mountain areas could be a consequence of the reforestation – natural and human-induced – of abandoned lands (Beguería *et al*., 2003; García-Ruíz *et al*., 2011; López-Moreno *et al*., 2011; Morán-Tejeda *et al*., 2011). No formal attribution was conducted in those studies, though, to assess the hydrological effects of land-cover changes. Those effects are quantified here in terms of the resulting changes in the average CN of each of the 43 watersheds included in the full reference dataset from 1956 to 1999 (Table 6). Note that CN has decreased in 33 out of the 43 watersheds analyzed within Guadalquivir River Basin. The maximum decline in CN was ca. 6% and occurred in station 6012. The largest increase occurred in station 5057, a small watershed of ca. 50 km$^2$ and was close to 16%. This though appears an atypical example. The area average CN for the 43 watersheds has decreased approximately -2%. The significance of those changes in CN is analyzed next.

*Hydrologic modelling* - Being inversely proportional to CN, the maximum soil water content, $H_{max}$, will increase as a result of a decline in CN, and consequently, the threshold precipitation value, $P_0$, will also increase (see the conceptual description of the model and Appendix I). Hence, in the absence of other changes in hydrological forcing, one would expect a decline in both surplus and surface runoff as a result of the historical changes in CN. The relative effects of these changes in CN compared to changes in hydrometeorological forcing was quantified through simulations of monthly streamflow conducted in two example watersheds ('Tranco de Beas', 6001 and 'Cala', 6014) with spatially averaged CN values that changed in time as estimated from the historical land-use analysis (see, for example, Fig. 9). These two example watersheds were used to represent contrasting hydrologic responses existing within the Guadalquivir basin, with the largest (positive) and lowest (most negative) correlation with PC2($Q$) (see Table 3). They are located one in then SE and the other in the NW of the basin, where distinct tendencies in the hydrological forcing variables have been identified (see station-by-station analysis above). Furthermore, the correlation of the observed streamflow series in the example watersheds with PC1($Q$) was high (see Table 3), hence, their behavior can be assumed as representative of the regional scale hydrologic response. The coefficient of determination between observed and simulated values in all cases ($R^2$ = 0.84 for station 6001, and $R^2$ = 0.76 for 6014) was high and comparable to other studies (Martos-Rosillo, *et al*., 2006; Carballo, *et al*., 2009). The simulated streamflows, in all the example watersheds follow closely the observations, with similar magnitudes and timing of peaks and base-flows (see Fig. 10a, b). In the two example stations, the simulated annual streamflows exhibited similar trends as the observed records (see Table 7 and compare with Table 5). In station 6001, in particular, annual streamflows in the simulations exhibit a significant decline of ca. 1.52 % year$^{-1}$, similar to the observed decline (1.59 % year$^{-1}$, also significant). The decreasing trends in the simulated streamflows were also significant, as in the observations, when analyzed season-by-season in winter, spring and summer, but not in fall. In station 6014, the annual streamflow trends are not significant neither in the simulations nor in the observations. Only in spring those trends are significantly negative, and this is consistent, also, between simulations and observations.

Average seasonal changes experienced by the different buckets and fluxes considered in the model are shown in Fig. 10c, d. Soil moisture content is almost nil in summer time, increasing in fall, and reaching near-saturation values in winter. It is at that time that evaporative fluxes are minimal and most of the precipitation becomes surplus. Only 30% of the surplus in 6001, becomes direct surface streamflow. The remaining 70% feeds the groundwater bucket and becomes streamflow with some delay. Groundwater becomes the only contribution to streamflow in summer time and early fall, when precipitation is largely used to reestablish the water content in the soil bucket. Precipitation increases in fall to reach maximal values from December to February, and is minimal in summer time. Evaporation rates increases from the lowest values in late fall and winter, peaking in late spring and early summer, and declining, thereafter, as a result of the declining soil moisture in summer time. On average, more than half of the annual rainfall (ca. 65% both in stations 6001 and 6014) is lost to the atmosphere through evaporation.

*Sensitivity analysis -* In the sensitivity exercise, the changes in streamflow resulting from perturbations in the average $H_{max}$ of the basin ($H_{max}$), the annual rainfall and evapotranspiration ($P_{ref}$ and $ET_0^{ref}$) or the seasonal rainfall distribution were assessed. The sign of the perturbations are those encountered in the analysis of the records: negative in $P$ and $H_{max}$, and positive in $ET_0$. To perturb the annual precipitation or the annual evapotranspiration, all monthly values were decreased/increased by a given and fixed percentage (5%). The perturbations in the seasonal distribution of precipitation were accounted for as follows. First, winter precipitation $P_w$ was decreased 5% while keeping the annual value constant. The amount of precipitation not falling in winter $\Delta P_w$ was uniformly reallocated to the rest of the year, so that the amount of precipitation falling in each of the other three seasons increased $\Delta P_w/3$. Similar experiments were then conducted by decreasing fall and spring precipitation ($P_f$ and $P_s$) in the same amount $\Delta P_w$. Summer precipitation was not perturbed, given that it is the season with the lowest, and almost nil, rainfall. The sensitivity coefficients for all factors considered in this exercise are shown in Table 8. Note first, that streamflows are most sensitive to changes in annual precipitation in both watersheds. The sensitivity to changes in the seasonal distribution of precipitation varies, though, depending on the watershed considered. In the case of station 6001 the annual streamflow is particularly sensitive to decreases in winter precipitation compared to reductions in other seasons. In station 6014, annual streamflow is slightly more sensitive to reductions in fall precipitation compared to the same reduction in winter precipitation (compare the sensitivity coefficients of $P_w$ and $P_f$). A reduction in spring precipitation compensated by increases in fall and winter even causes streamflows to increase in both cases. The changes in $ET_0$ and $H_{max}$ lead to a decrease in annual streamflow and have similar sensitivity coefficients.

The relative contribution of climatic (changes in the annual precipitation and evapotranspiration, and in the seasonal distribution of precipitation) and non-climatic factors (land-use changes) on the variability of the streamflow records used in our analysis was assessed from Eq. 3, using the coefficients of variation and the sensitivity coefficients shown in Table 8. A relative change in the Curve Number of 5.13% ($\Delta H_{max}/H_{max-0} = 0.0513$), as estimated, for example, for 6001 from 1950 to 2005, will lead to changes in streamflows of up to -2.8%. A similar decrease in annual precipitation of 7.4%, has occurred in the study period, but this change could lead to a reduction of more than 15% in both watersheds. These changes in precipitation, though, have been shown to be statistically non-significant. The effect of the reduction in annual $ET_0$ is very similar in both stations, where a decrease in the annual streamflow of approximately 3.5% has been calculated. The largest and most significant changes in the last 56 years, though, has in the seasonal rainfall distribution. Precipitation has significantly decreased in almost 38% during winter time; it has increased, but not significantly, in fall (+12.9%), and decreased also not significantly in spring (-1%). The redistribution of winter precipitation could drive changes of almost 23% in both stations, being the highest contribution to the variability of the annual streamflow. According to this modelling exercise, and if we accept that Eq. 3 is valid, the streamflows should have decreased in ca 30% during the last 56 years in both stations. Almost 99% of that change can be attributed to climate factors, being the remaining 1% attributed to land-use changes.

*Grid of synthetic streamflow data* - The time series of *P* and $ET_0$ at every grid point in the Spain02 dataset was used to generate time series of monthly streamflow, $Q_g$. The calibrated model for station 6001 were used in these simulation exercises (similar results are obtained for the other stations). Monthly values of all model variables (including moisture deficit, groundwater reserves, ...) were also stored in each grid point for each simulation. In 118 out of the 192 series (61%) the trends in annual precipitation were negative; in only 27 those trends were significant. In 145 (almost 75%) of the 192 synthetic streamflow series the annual values had negative trend; in 24 those trends were also significant at 95% confidence level. Only 3 of the 192 streamflow series had a significant positive trend. This suggests, first, that even if precipitation does not exhibit any significant trends, streamflows may change significantly. Note also, that the majority of the streamflows in the Guadalquivir basin should exhibit negative trends.

Using average values of the model variables for each season, we re-constructed and plotted the curve used to convert precipitation $P$ into surplus $S$ (see Fig. 11a). Note that winter curve is steeper for the range of monthly precipitation values below its 95 percentile (Fig. 11b, c). This is partly as a result of the lower soil moisture deficit (hence, lower precipitation thresholds, $P_0$) and the lower $ET_0$ (hence, lower $\delta$) in winter compared to other seasons. Several consequences arise from the differences in model curves among seasons. First, the largest surplus for any given and fixed reference value of precipitation $P'$ will occur in winter. Also, the largest changes in streamflow occurring in response to a perturbation of any given magnitude in seasonal precipitation $\Delta P'$ from that reference value $P'$ will occur in winter. Hence, a reduction in -$\Delta P'$ in winter precipitation, will cause a reduction -$\Delta Q$ in annual streamflow, which exceeds the possible increase in streamflow resulting from an increase in precipitation of similar magnitude +$\Delta P'$ occurring in any of the other seasons.

## 4. Conclusions

Detecting, attributing and accounting for long-term climate-driven trends in streamflows is probably one of the major challenges faced by water managers in the semi-arid Mediterranean Basin during the next decades. Trend detection and attribution, is not a straightforward task and needs to be addressed carefully and at the basin scale. In this work, streamflow records from a set of watersheds with minimal anthropic influence in the Guadalquivir River Basin, the southern-most of the larger watersheds in Spain facing recurrent structural deficits in water resources, have been analyzed to determine whether streamflows have changed or not during the last 50 years as a result of climatic changes, and to understand the dominant drivers of those changes. Regional scale streamflows in the Guadalquivir Basin are subject to strong seasonal variations which are largely associated to the seasonal distribution of rainfall, with peak values in winter and the lowest values in summer time. Potential evapotranspiration also exhibits large variations, being largest in summer time. The conclusions arrived at in this analysis can be summarized as follows:

[1] Regional scale streamflows have undergone a significant reduction (of ca. 60%) during the last five decades, especially noticeable since 1970's. The reduction in streamflow volumes are detected not only when the analysis is done at annual scales, but also, when done on a season-by-season basis. These results of the analysis conducted on the regional signal are consistent with those obtained when analyzing one-by-one on a set of gauging points, and with the results of calibrated lumped-parameter rainfall-runoff model simulating streamflow from two individual watersheds in the Guadalquivir Basin. Assuming that the trends in streamflow and the pressure on surface water resources (75% of the total 4000 hm$^3$/year used in the basin, Berbel *et al.*, 2012) remain constant in the next 50 years, given that surface water resources available today in the basin have been recently evaluated and sum up to 7043 hm$^3$/year (Berbel *et al.*, 2012) water managers will be working under a continuous structural water deficit scenario by the end of the century.

[2] Trends in hydrological forcing and land-use were examined to identify the driving factors of streamflow changes in the reference hydrologic network. Annual rainfall have not changed significantly, in a statistical sense, during the study period. Significant changes, though, are observed in the intra-annual distribution of precipitation: the largest changes have occurred in winter, when precipitation exhibits significantly

decreasing trends; in spring, precipitation has also decreased, but not significantly; in summer and fall, precipitation has increased, though not significantly either. As for $ET_0$, there has been a generalized and significant increase throughout the region, annually and particularly for the winter and spring seasons. As a result of the increasing $ET_0$ and the redistribution of precipitation from seasons with low $ET_0$ (winter) to others with large $ET_0$ (summer and fall) one should expected reductions in streamflow. Finally, analyzing of a series of land-use maps from 1956 to 1999, we find that the surface of woods and forest, representing more than 60% of the area in the study watersheds, has increased since 1956, leading to reductions in hydrological Curve Number and consequently to reductions in streamflow.

[3] A calibrated lumped-parameter physically based continuous model is used, in a sensitivity analysis to assess the effects of climatic and non-climatic factors as drivers of streamflow changes. The largest sensitivity of streamflows to winter precipitation compared to the precipitation in other seasons is the result of steeper rainfall-runoff curves, arising as a consequence of lower soil moisture deficit and the lower $ET_0$, prevailing during the coldest months of the year. The results of the sensitivity analysis are used in an attribution exercise, to understand the long-term changes detected in our streamflow records. In this attribution exercise, the annual changes in precipitation are discarded as a plausible cause of streamflow change given that they are statistically non-significant. The observed changes in annual streamflows can be explained, in the model, as a result of the statistically significant changes in $ET_0$, and winter precipitation. In the model attribution analysis, land-use changes are shown to lead to small changes in streamflow, when compared with the effects of changes in $ET_0$ or the seasonal redistribution of precipitation.

## 5. Acknowledgements

The first author was supported by the Ministry of Education, Culture and Sport of Spain (*Beca de colaboración* 2014). This work is part of the Master's thesis presented by the second author. He was supported by the University of Granada (*Plan Propio de Investigación* 2010) while working on this manuscript under the supervision of the last author. This work was partially funded by the Spanish Ministry of Science and Innovation through projects CGL2008-06101, CGL2008-05016, CGL2013-48539-R with additional support from the European Community Funds (FEDER) and P11-RNM-7941 (Junta de Andalucía-Spain). We want to thank *Centro de Estudios y Experimentación de Obras Públicas,* CEDEX, and *Confederación Hidrográfica del Guadalquivir* CHG, for making the data available to us.

Zhang, Y., Shao, Q., Xia, J., Bunn Stuart, E., Zuo, Q., 2011. Changes of flow regimes and precipitation in Huai River Basin in the last half century. Hydrological processes. 25, 246-257.
43

Appendix I. Equations in the hydrological model proposed by Temez.

| | |
|---|---|
| Beta | $\beta = e^{-\alpha D/2}$ |
| Surplus | $S_i = \dfrac{(P_i - P_{0i})^2}{P_i + \delta_i - 2P_{0i}}$ |
| Precipitation threshold | $P_{0i} = C(H_{max} - H_{i+1})$ |
| Deep percolation | $G_i = I_{max} \dfrac{S_i}{S_i + I_{max}}$ |
| Groundwater flow | $g_i = g_{0i} \cdot e^{-\alpha \cdot t}$ |
| Groundwater reserves | $R_i = \int_t^\infty g_i\,dt = \dfrac{g_i}{\alpha}$ |
| Groundwater contribution to streamflow | $A_{Fi} = \left(\dfrac{g_{i-1}}{\alpha} + G_i\right) - \left(G_i + \dfrac{g_{i-1} \cdot e^{-\alpha D_i/2}}{\alpha}\right)e^{-\alpha D_i/2}$ <br><br> $A_{Fi} = \dfrac{g_{i-1}}{\alpha}(1 - e^{-\alpha D_i}) + G_i(1 - e^{-\alpha D_i/2})$ <br><br> $A_{Fi} = A_{min} + (1-\beta)G_i$ |
| Minimum streamflow | $A_{min} = \dfrac{g_{i-1}}{\alpha}(1 - \beta^2)$ |
| Total streamflow | $A_i = (S_i - G_i) + (A_{min} + (1-\beta)G_i)$ <br><br> $A_i = A_{min} + S_i - \beta G_i$ |
| Water available during an interval | $X_i = H_{i-1} + P_i - S_i$ |
| Soil moisture in the next interval | $H_{i+1} = X_i - ETP_i \;(if\, X_i \geq ETP_i)$ <br> $H_{i+1} = 0 \;(if\, X_i \leq ETP_i)$ |

**Figure Captions**

Figure 1. (a) Guadalquivir Basin in southern Spain, with the reference watersheds identified. The subset of reference gauging stations used in the long-term analysis are shown with solid circles, and their reference number is given. (b) Permeability map of the Guadalquivir Basin.

Figure 2. Results of the regional analysis. (a) Loading factors of the first Empirical Orthogonal Function for Precipitation, EOF1(P); (f) Time series of the first Principal Component of Precipitation, PC1(P) with its temporal trend and its significance; (k) Seasonal changes represented in the first principal component of precipitation PC1(P). (b)-(g)-(l): as (a)-(f)-(k), but for the second EOF/PC of precipitation. (c)-(h)-(m), for first EOF/PC of $ET_0$. (d)-(i)-(n): for first EOF/PC of streamflow. (e)-(j)-(o): second EOF/PC of streamflow.

Figure 3. Time series of mean annual first (a) and second (b) principal component of streamflow, PC1(Q) and PC2(Q), resulting from the analysis conducted with records with varying length, and number of stations. The length of the time series analyzed are shown in the middle. The stations included in each case, are those with at most 20% of their data filled (Table 1).

Figure 4. Sen's slopes of precipitation and $ET_0$ records from observational data set (obtained from Spain02 grid). Filled symbols are used to indicate significance at the 90% confidence level. Solid borders indicate that trends are significant at the 95% confidence level.

Figure 5. Time series of the first principal component of rainfall PC1(*P*), streamflow PC1(Q), and $ET_0$, PC1($ET_0$), for each season: (a) fall; (b) winter; (c) spring; and, (d) summer.

Figure 6. Sequential version of Mann-Kendall test for the first principal component of rainfall PC1(*P*), streamflow PC1(Q), and $ET_0$, PC1($ET_0$), for each season: (a) fall; (b) winter; (c) spring; and, (d) summer.

Figure 7. Linear correlation of winter precipitation and winter streamflow (left) and linear correlation of fall precipitation and the residual streamflow of winter (right).

Figure 8. Standardized values of spring streamflow, spring precipitation and winter precipitation (from top to bottom). Vertical lines identify years with streamflow volumes during spring time that are large or small, compared with rainfall volumes in that same season.

Figure 9. Maps of Curve numbers for Tranco de Beas watershed (6001), corresponding to the land-use prevaling in (a) 1956 and in (b) 1999. In (c), we have plotted the difference in CN between 1956 and 1999. Blanked areas are those without changes in their land use.

Figure 10. Simulated (thick line) and observed (symbols with narrow lines in between) monthly streamflow during 20 years within the study period, for gauging station (a) 6001; and (b) 6011. Average monthly series of fluxes and storage volumes considered

in the hydrologic model are shown in subplots (c) and (d): precipitation (P), streamflow (Q), real evapotranspiration (E), infiltration (I) to groundwater storage (V), and soil moisture content (H). Results for watershed 6001 are shown in (c), and for 6014 in subplot (d).

Figure 11. (a) Rainfall-runoff curve used in the model to transform precipitation P in surplus S. The curve is zero, for $P < P_0$, and increases gradually to become linear and with a unit slope asymptotically. (b) Average S vs. P curves for station 6011, and for different seasons. (c) Same as (b) but for station 6011.

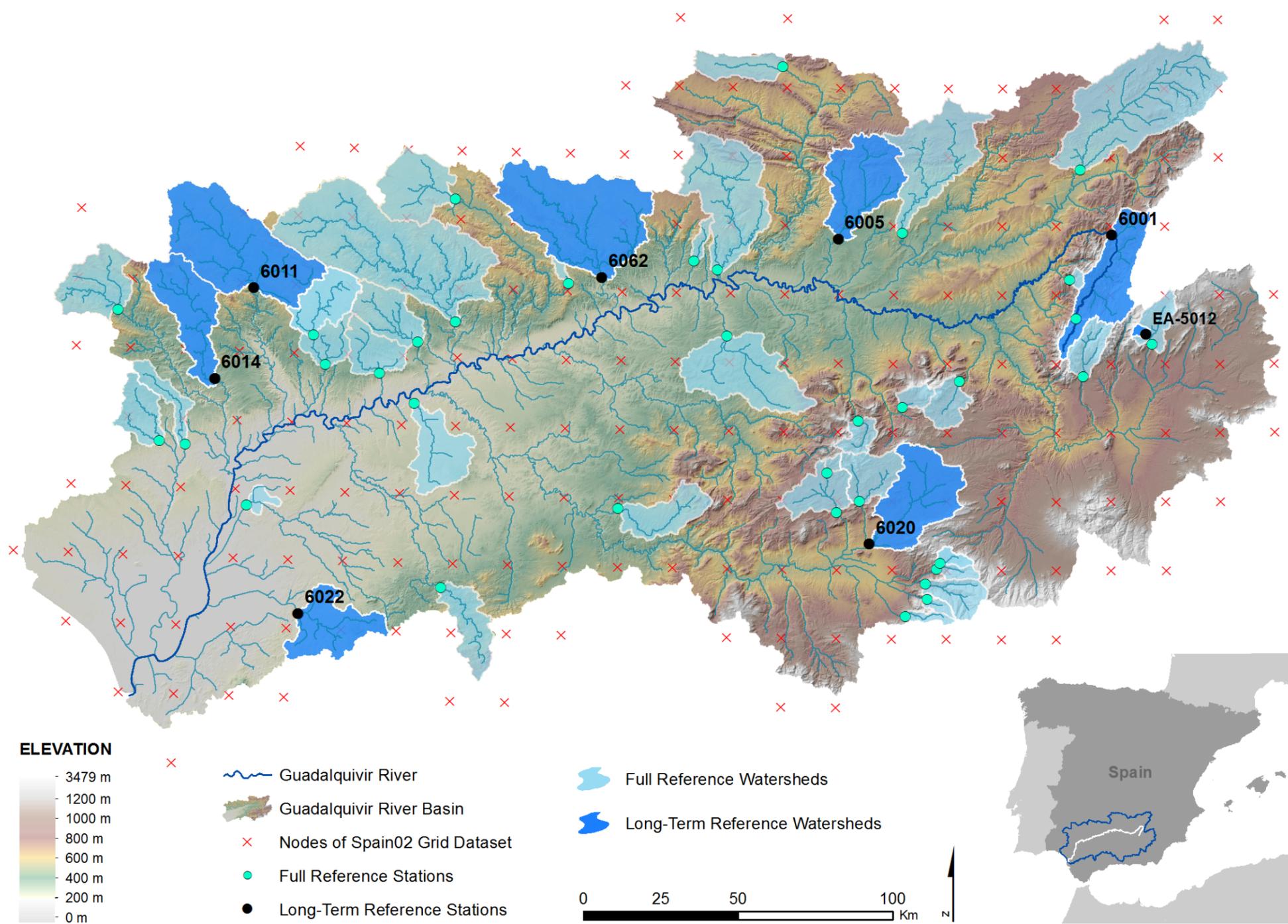

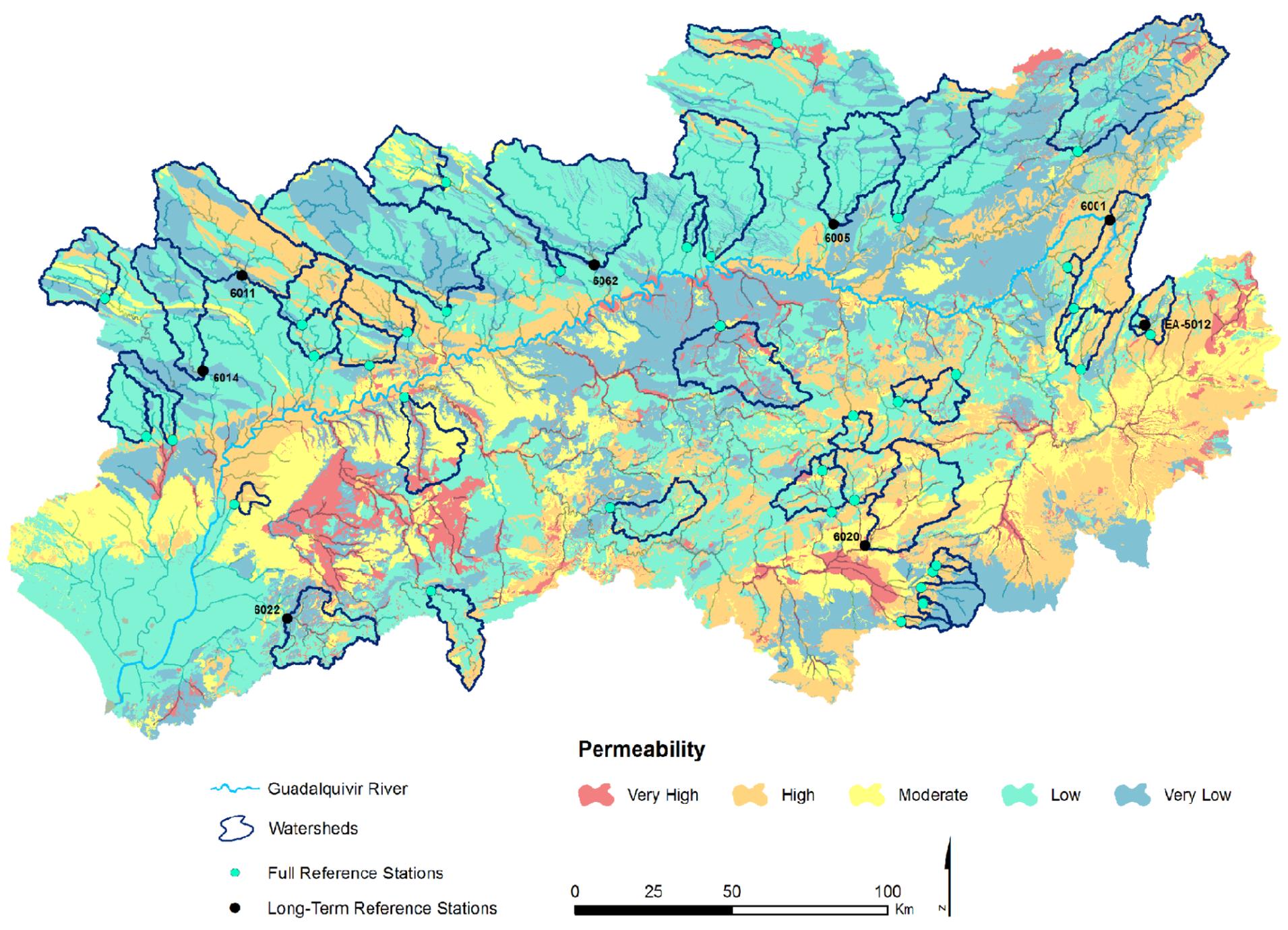

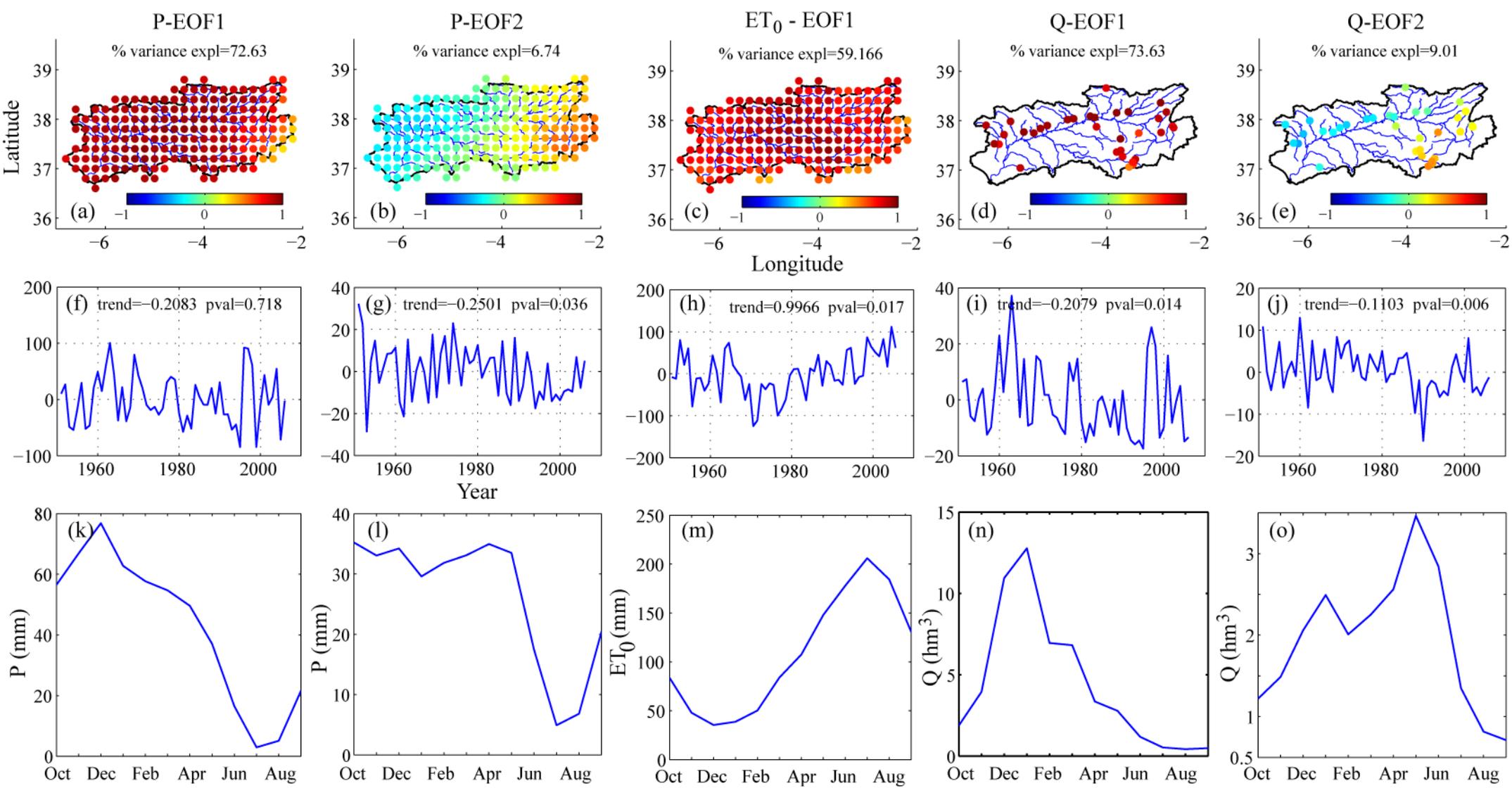

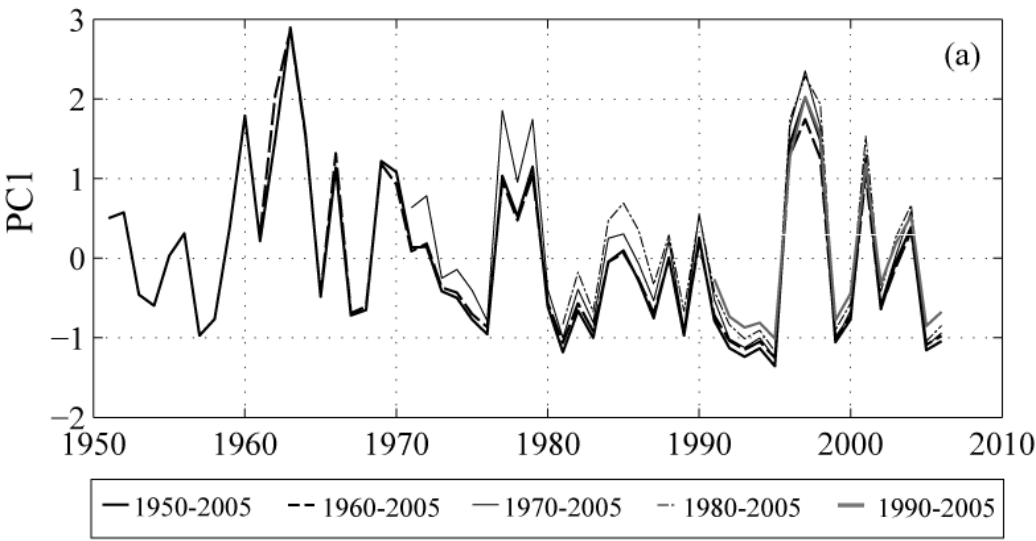
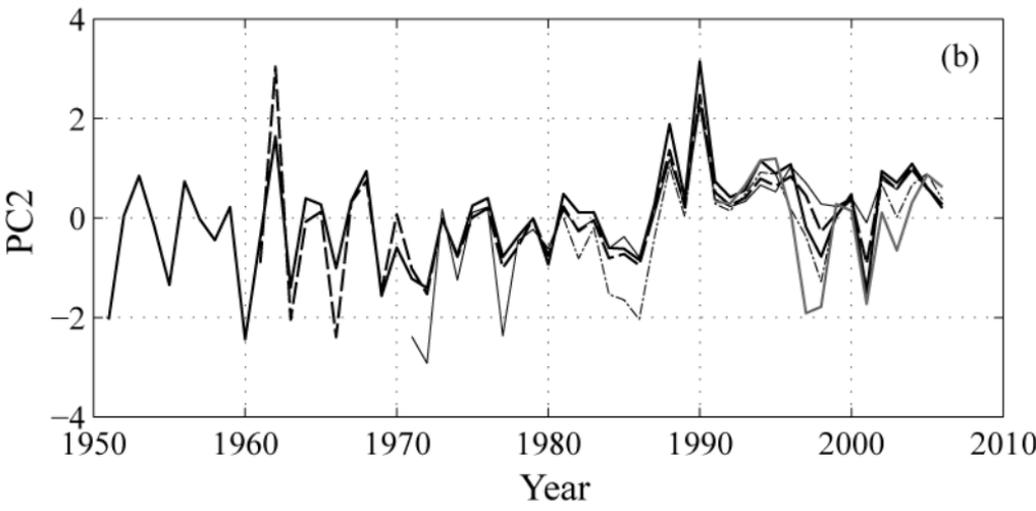

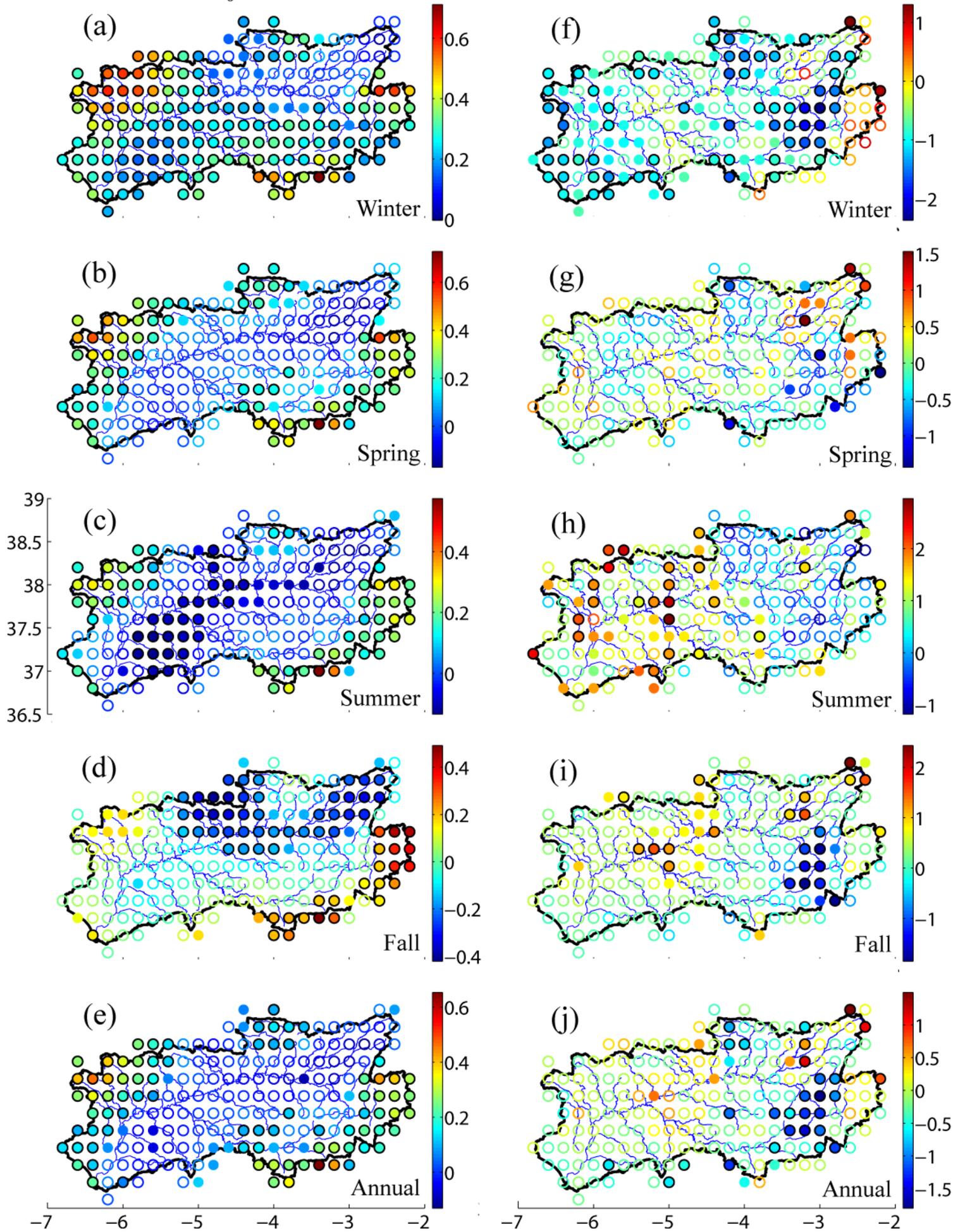

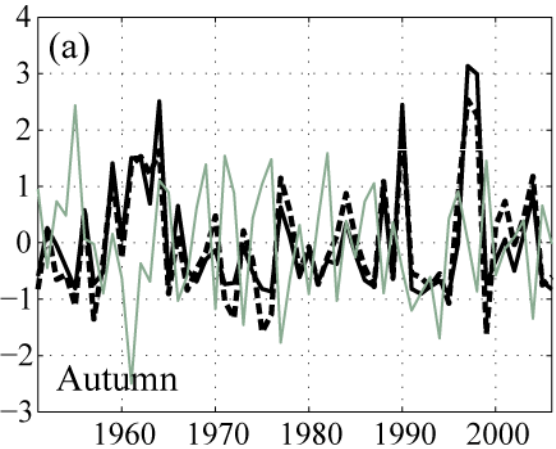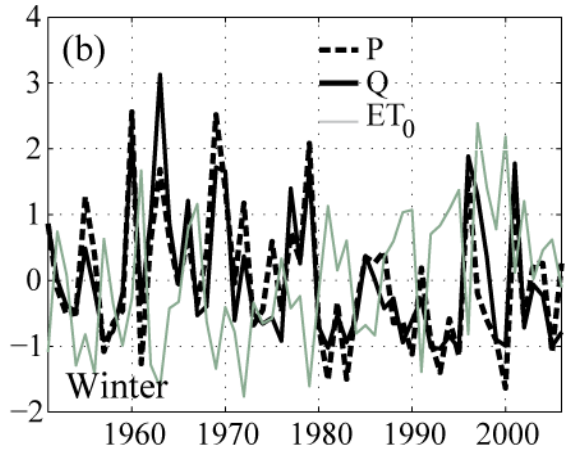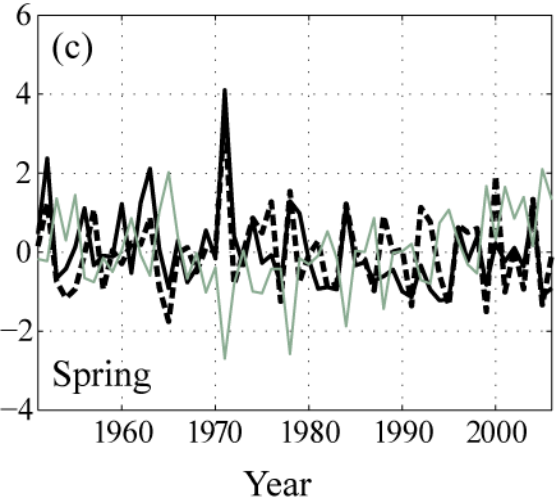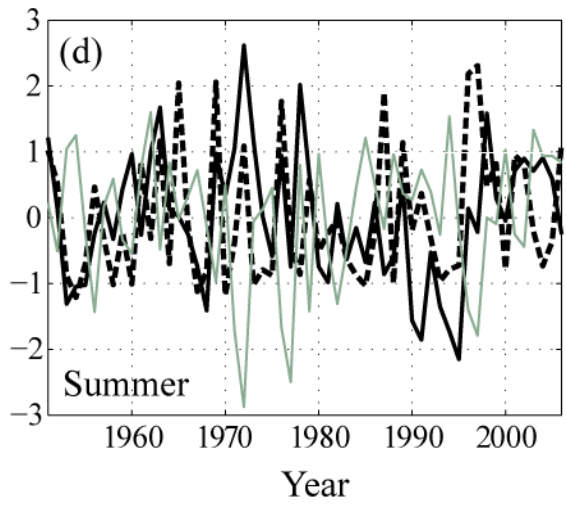

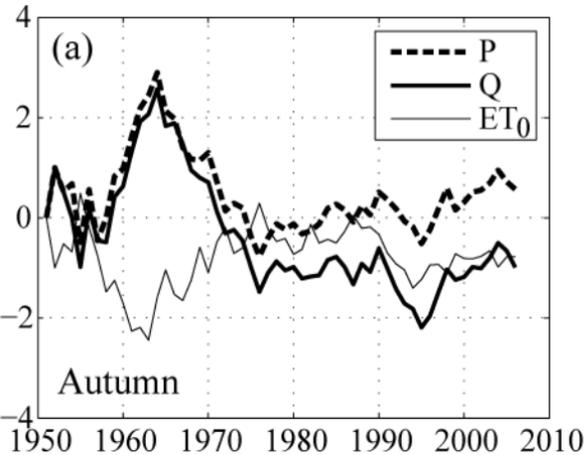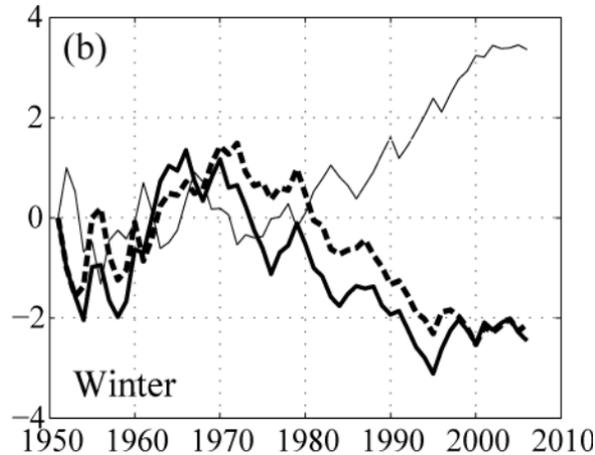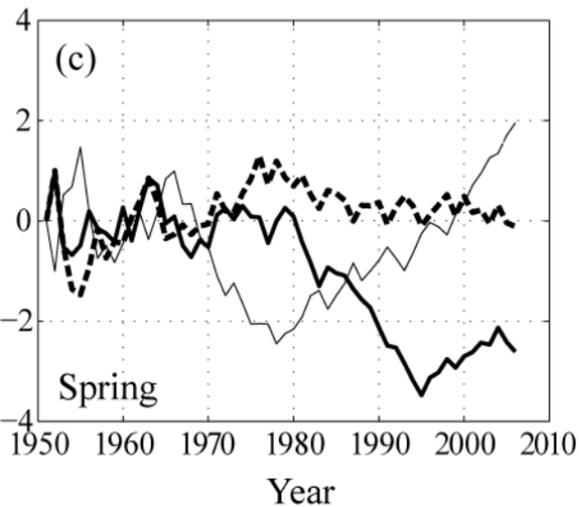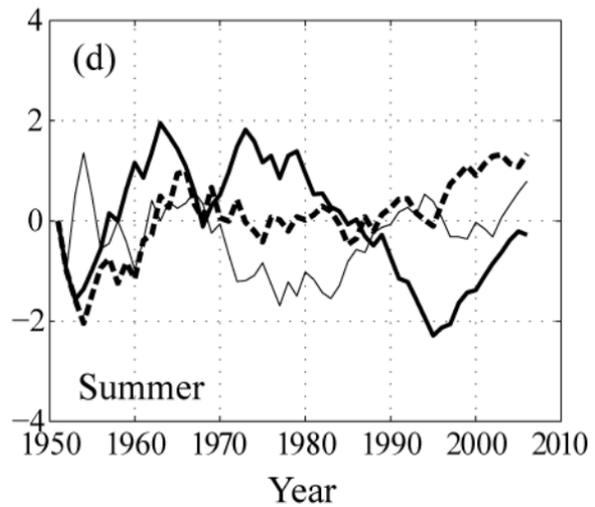

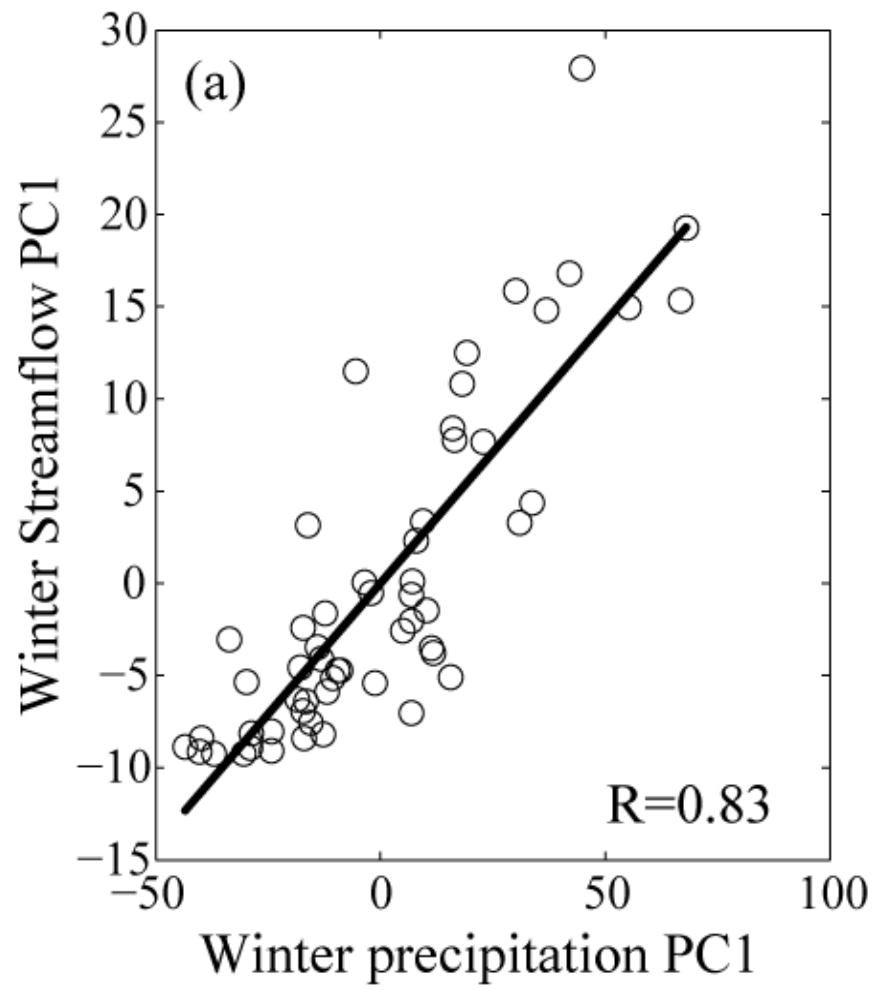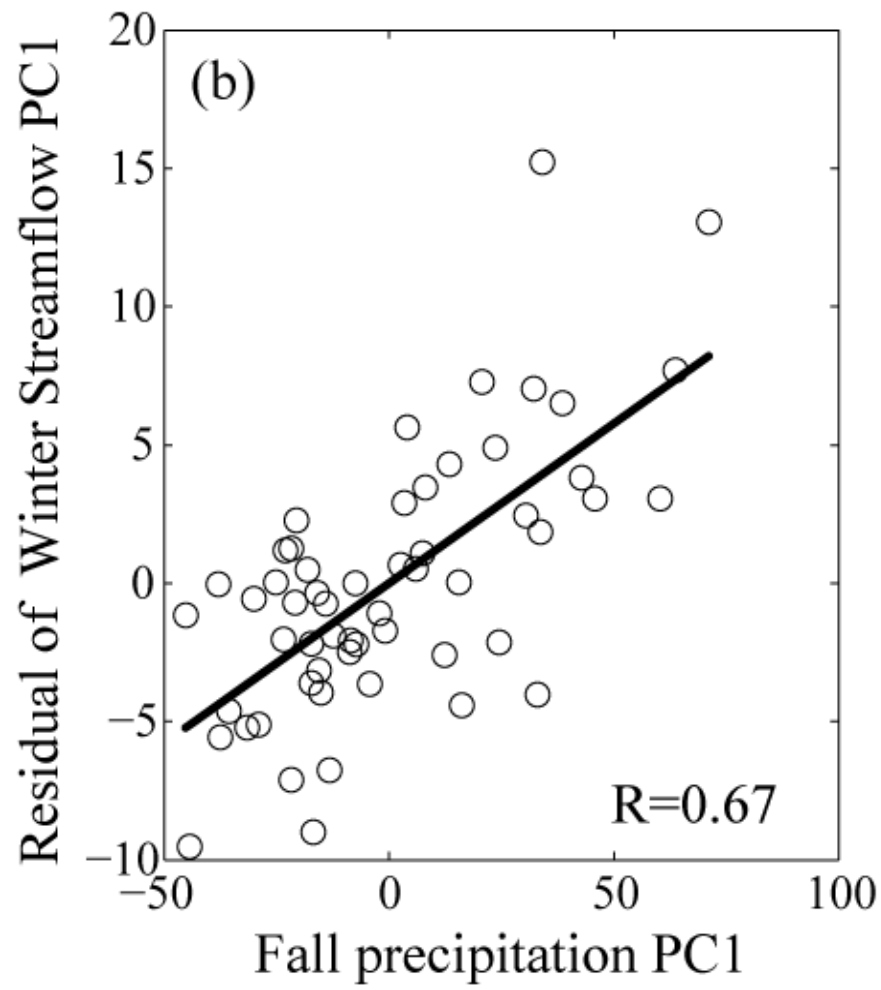

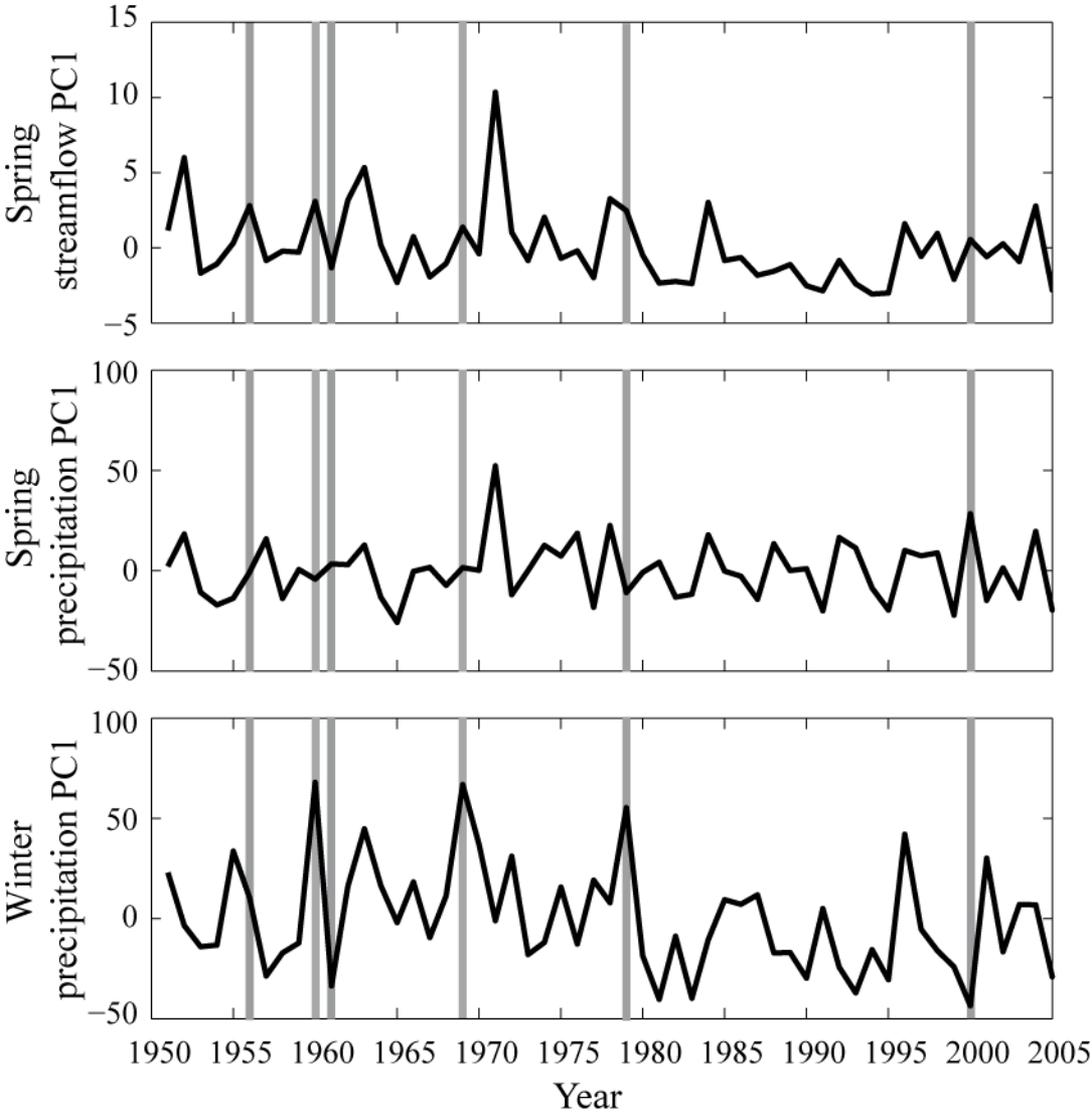

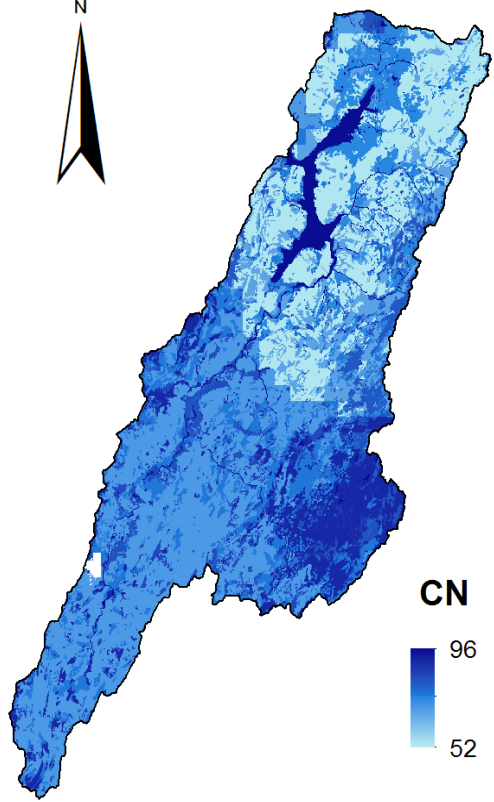
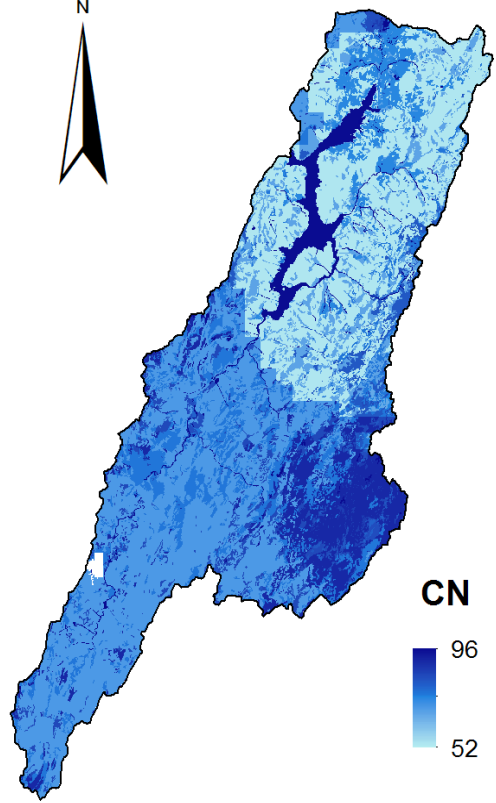
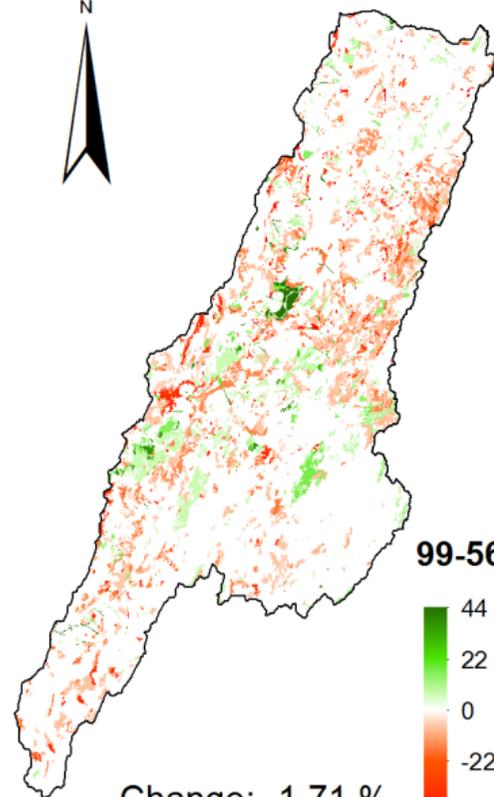

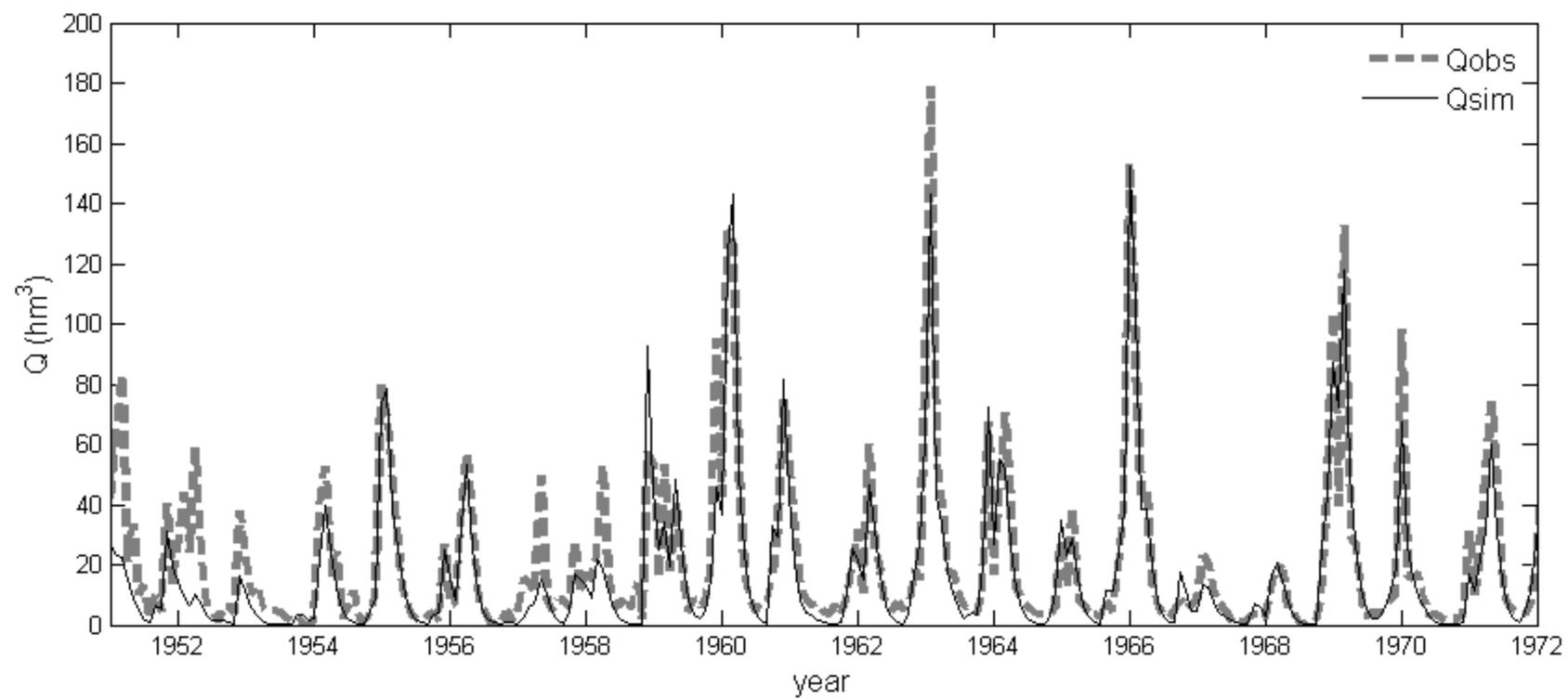
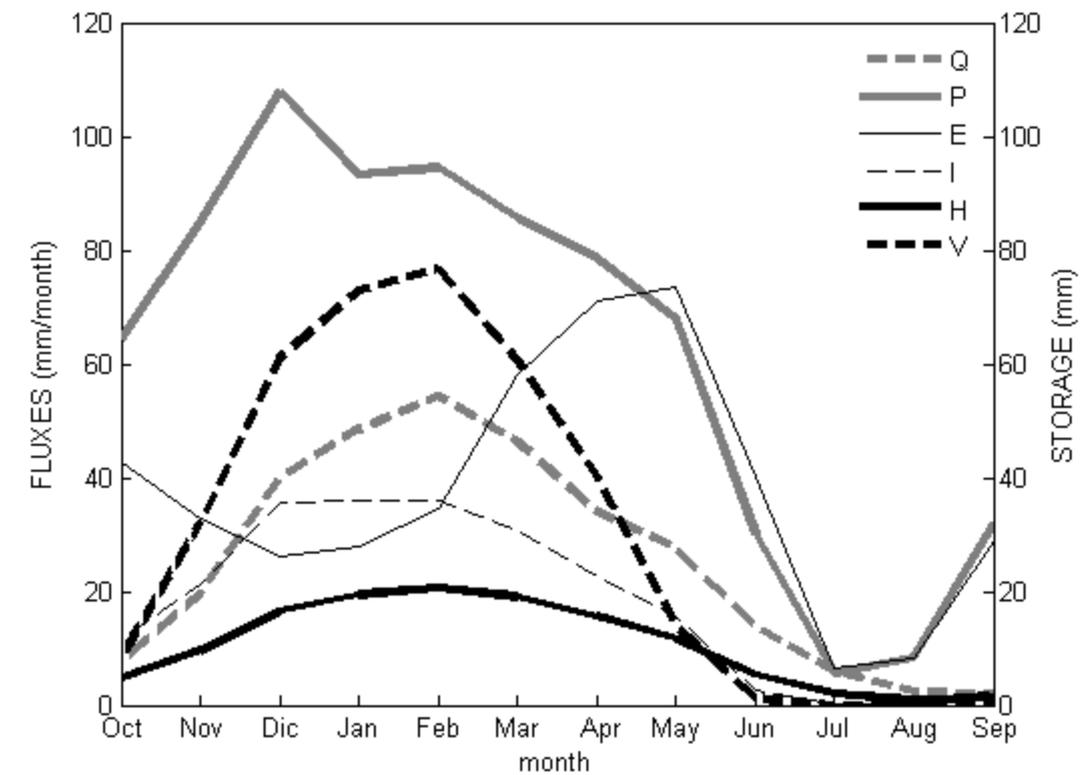
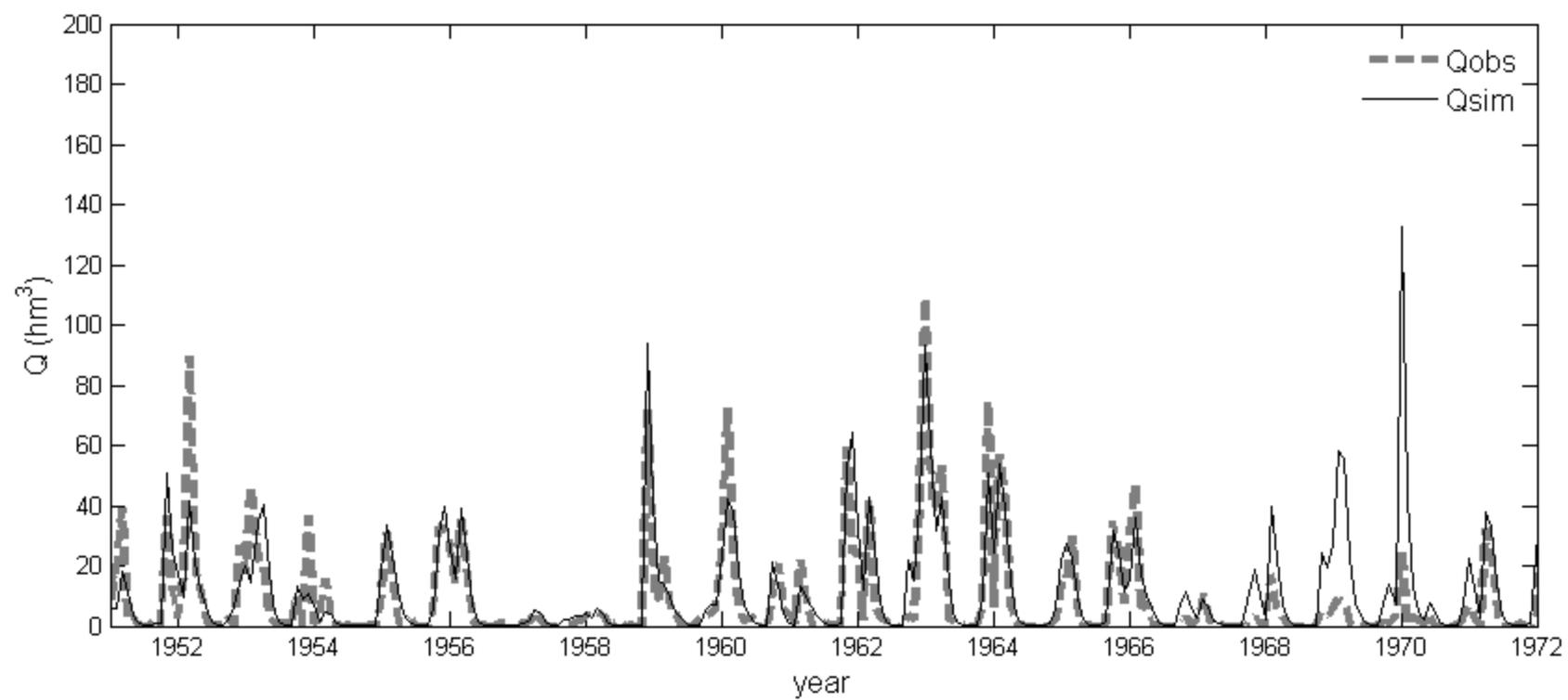
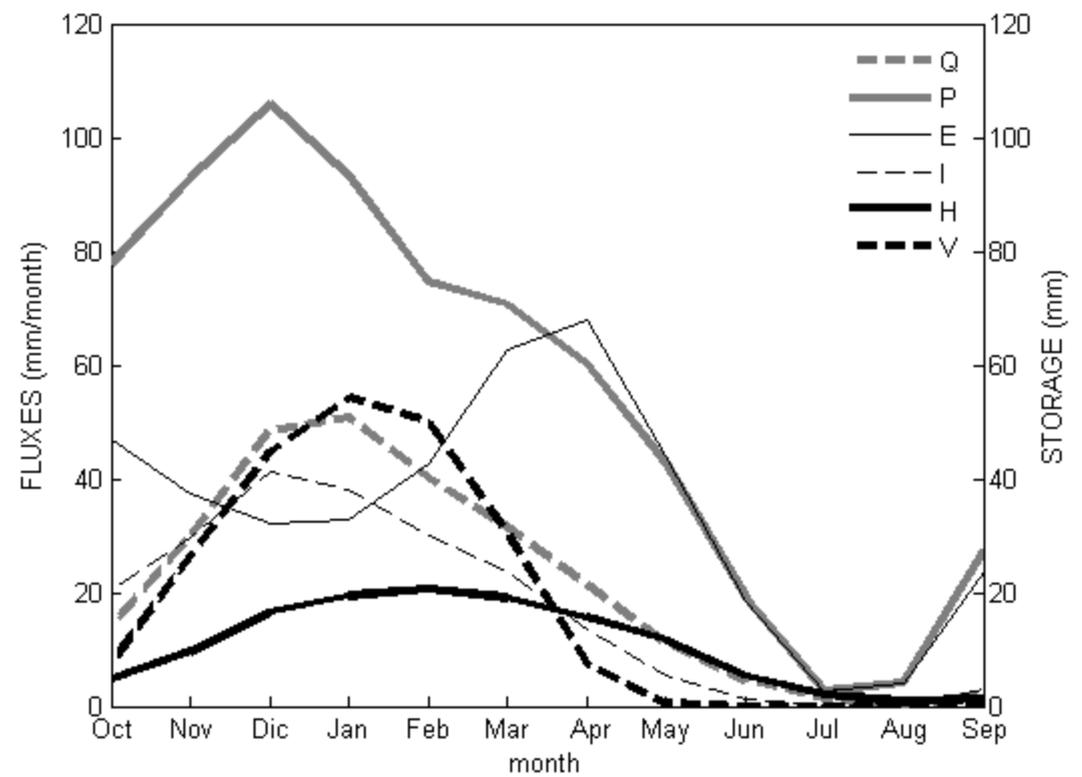

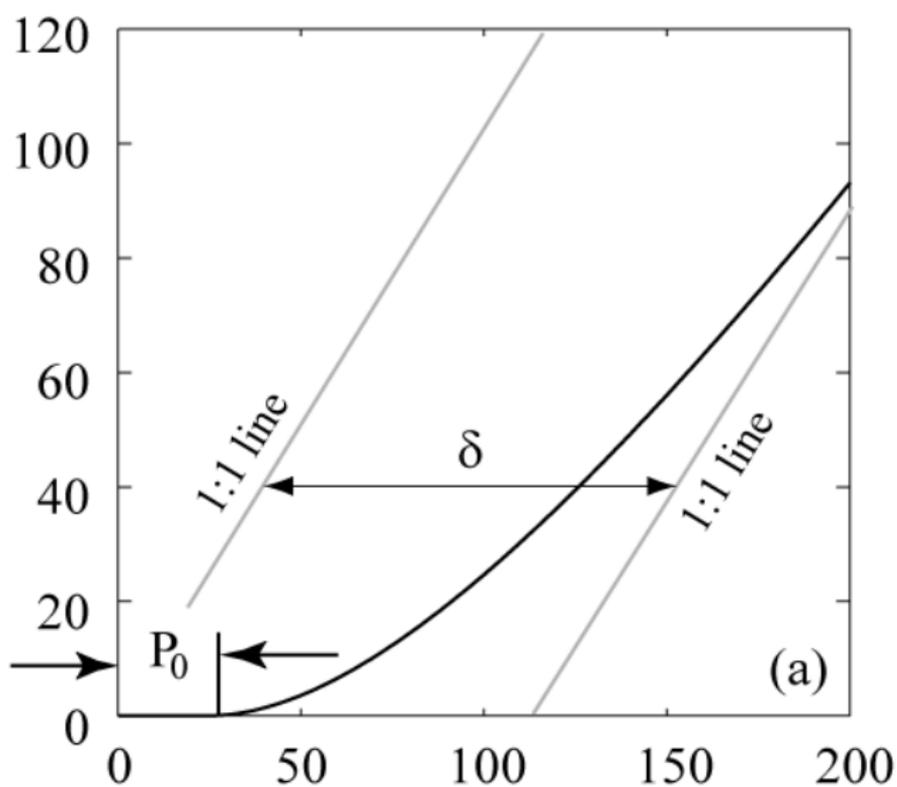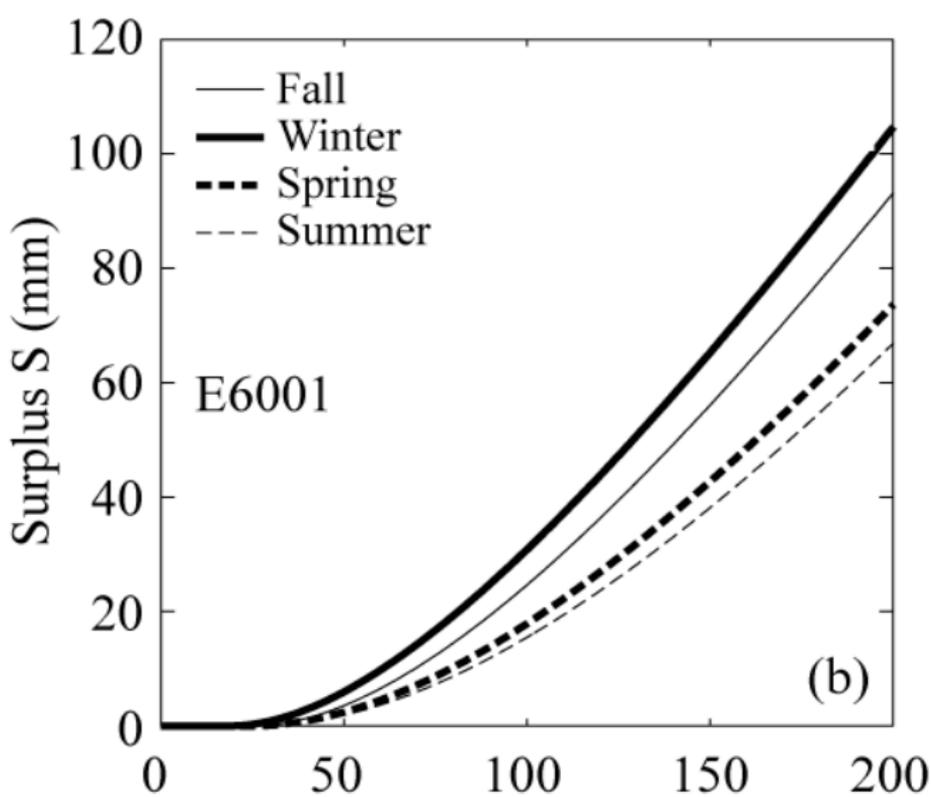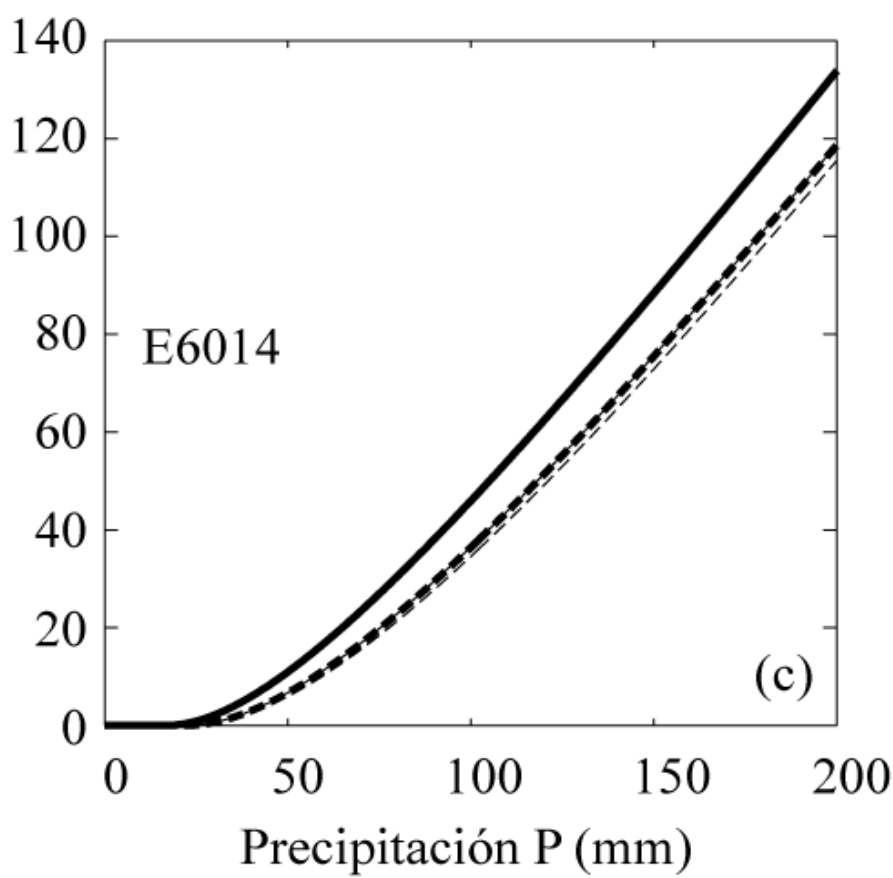

| Percentage of filled data | | < 5% | < 20% |
|---|---|---|---|
| **Sub-period** | 50-05 | 6 | 8 |
| | 55-05 | 7 | 8 |
| | 60-05 | 8 | 10 |
| | 65-05 | 8 | 13 |
| | 70-05 | 12 | 16 |
| | 75-05 | 13 | 15 |
| | 80-05 | 14 | 16 |
| | 85-05 | 14 | 23 |
| | 90-05 | 20 | 34 |
| | 95-05 | 30 | 43 |

Table 1. Number of stations included in the different periods considered with filled data thresholds of 5% and 20%.

Table 2. General characteristics of the stations comprising the "full reference data set".

| Code | Name | Surface (km$^2$) | Elevation (m) | P (mm) | ET$_0$ (mm) | Code | Name | Surface (km$^2$) | Elevation (m) | P (mm) | ET$_0$ (mm) |
|---|---|---|---|---|---|---|---|---|---|---|---|
| 5012 | EA-5012 El Doctor | 17.05 | 1475.71 | 518.4 | 1079.6 | 6017 | E39 Retortillo | 299.76 | 470.97 | 548.8 | 1293.6 |
| 5021 | EA-5021 Quentar | 46.95 | 1416.43 | 511.9 | 989.1 | 6018 | E15 Guadalmena | 1224.80 | 909.53 | 530.3 | 1129.2 |
| 5024 | EA-5024 Hornos del Vidrio | 156.30 | 1098.81 | 446.2 | 1140.1 | 6020 | E45 Cubillas | 635.62 | 1038.15 | 469.4 | 1144.0 |
| 5039 | EA-5039 Central Diechar | 35.41 | 2050.73 | 580.7 | 892.6 | 6022 | E68 Torre del Águila | 445.03 | 147.07 | 581.5 | 1328.6 |
| 5056 | EA-5056 Gerena | 153.63 | 288.88 | 659.6 | 1325.4 | 6029 | E05 Bolera | 163.86 | 1443.52 | 600.1 | 1078.6 |
| 5057 | EA-5057 Guadaira | 52.14 | 32.04 | 511.8 | 1411.6 | 6037 | E33 Sierra Boyera | 424.01 | 587.08 | 497.7 | 1242.9 |
| 5086 | EA-5086 Dilar | 45.21 | 2005.00 | 562.2 | 916.3 | 6038 | E42 Quéntar | 101.31 | 1533.87 | 507.5 | 987.7 |
| 5095 | EA-5095 Tozar | 247.76 | 965.84 | 558.2 | 1188.1 | 6039 | E17 Quiebrajano | 96.42 | 1210.35 | 639.7 | 1131.6 |
| 5097 | EA-5097 Los Piedros | 297.24 | 560.16 | 525.4 | 1258.9 | 6045 | E02 Aguascebas | 22.03 | 1332.66 | 718.0 | 1140.7 |
| 5123 | EA-5123 Madrefuentes | 371.59 | 158.30 | 457.2 | 1367.3 | 6046 | E35 Guadañuno | 23.54 | 578.61 | 650.6 | 1316.9 |
| 5134 | EA-134 Cerrada del Utrero | 58.54 | 1371.00 | 644.5 | 1110.4 | 6047 | E27 Martín Gonzalo | 48.36 | 545.73 | 622.8 | 1266.2 |
| 5137 | EA-5137 Cambil | 67.46 | 1287.16 | 474.0 | 1175.2 | 6048 | E41 Canales | 176.26 | 1974.73 | 518.6 | 953.0 |
| 5138 | EA-5138 Puertollano | 212.15 | 741.11 | 444.7 | 1180.4 | 6049 | E26 Yeguas | 799.38 | 649.17 | 560.6 | 1232.4 |
| 5142 | EA-5142 Salado de Porcuna | 659.32 | 417.94 | 488.4 | 1328.4 | 6050 | E44 Colomera | 241.80 | 1077.93 | 553.1 | 1165.0 |
| 5145 | EA-5145 El Pedroso | 294.89 | 577.74 | 609.5 | 1243.0 | 6052 | E14 Fernandina | 658.11 | 748.57 | 446.6 | 1211.4 |
| 5152 | EA-5152 Frailes | 22.37 | 1207.61 | 654.9 | 1146.7 | 6055 | E54 José Torán | 237.74 | 362.44 | 581.4 | 1321.7 |
| 6001 | E01 Tranco de Beas | 560.69 | 1191.65 | 752.5 | 1029.1 | 6056 | E56 Huesna | 469.84 | 519.61 | 617.9 | 1267.2 |
| 6005 | E19 Rumblar | 577.02 | 693.31 | 475.2 | 1228.3 | 6058 | E67 Agrio | 231.07 | 301.67 | 785.5 | 1315.1 |
| 6011 | E57 El Pintado | 1138.09 | 552.49 | 592.6 | 1210.1 | 6060 | E03 San Clemente | 155.14 | 1390.99 | 517.0 | 1045.0 |
| 6012 | E37 Bembézar | 1635.99 | 499.51 | 531.0 | 1255.8 | 6061 | E55 Puebla de Cazalla | 294.92 | 551.35 | 618.2 | 1193.3 |
| 6014 | E64 Cala | 525.39 | 506.30 | 670.6 | 1231.6 | 6062 | E29 Guadalmellato | 1201.42 | 594.79 | 544.6 | 1268.4 |
| 6016 | E61 Aracena | 404.20 | 558.80 | 764.2 | 1187.7 | — | — | — | — | — | — |

| Station ID | Longitude | Latitude | PC1(Q) | PC2(Q) |
|---|---|---|---|---|
| 5012 | -2.67 | 37.89 | 0.76 | -0.37 |
| 6001 | -02.8 | 38.17 | 0.78 | **-0.50** |
| 6005 | -3.81 | 38.16 | 0.91 | -0.09 |
| 6011 | -5.96 | 37.98 | 0.88 | 0.40 |
| 6014 | -6.09 | 37.72 | 0.81 | 0.44 |
| 6020 | -3.68 | 37.27 | 0.69 | -0.41 |
| 6022 | -5.76 | 37.04 | 0.87 | 0.12 |
| 6062 | -4.68 | 38.04 | 0.91 | 0.26 |

Table 3. Loadings of the eight gauging stations used for long-term reference. with the PC1(Q) and PC2(Q).

Table 4: Trends (in percentage/year) for the reconstructed series associated PCs series of streamflow precipitation and potential evapotranspiration. Significant trends at 95% confidence level are in bold.

|  | Winter | Spring | Summer | Fall | Annual |
|---|---|---|---|---|---|
|  | % year$^{-1}$ | % year$^{-1}$ | % year$^{-1}$ | % year$^{-1}$ | % year$^{-1}$ |
| PC1($Q$) | **-1,317** | **-1,125** | 0,539 | -0,186 | *-0,921* |
| PC2($Q$) | -0,128 | **-0,546** | 0,469 | -0,158 | *-0,218* |
| PC1($P$) | **-0,601** | -0,184 | 0,567 | 0,073 | -0,139 |
| PC1(E$T_0$) | **+0.270** | **+0.101** | +0.020 | -0.044 | **+0.067** |

|            |           |          | Streamflow (Q) |        |        |        |        |
| ---------- | --------- | -------- | -------------- | ------ | ------ | ------ | ------ |
| *Station ID* | *Longitude* | *Latitude* | *Winter* | *Spring* | *Summer* | *Autumn* | *Annual* |
| 5012 | -2.67 | 37.89 | **-1.08** | -0.75 | 0.28 | -0.83 | *-0.85* |
| 6001 | -2.80 | 38.17 | **-1.89** | **-1.35** | **-1.91** | -0.90 | **-1.59** |
| 6005 | -3.81 | 38.16 | -1.31 | -0.93 | **-2.38** | -0.53 | *-1.69* |
| 6011 | -5.96 | 37.98 | *-1.20* | **-1.64** | **-3.92** | -0.57 | -1.16 |
| 6014 | -6.09 | 37.72 | *-1.75* | **-2.33** | 0.00 | 0.09 | -0.96 |
| 6020 | -3.68 | 37.27 | **-2.10** | **-2.55** | 0.30 | **-1.61** | **-1.88** |
| 6022 | -5.76 | 37.04 | -0.92 | -1.47 | 0.00 | 0.00 | -0.72 |
| 6062 | -4.68 | 38.04 | -1.12 | **-1.92** | 0.00 | -0.66 | *-1.50* |
|            |           |          | Precipitation (P) |        |        |        |        |
|            |           |          | *Winter* | *Spring* | *Summer* | *Autumn* | *Annual* |
| 5012 | -2.67 | 37.89 | -0.71 | 0.22 | 0.34 | -0.23 | -0.28 |
| 6001 | -2.80 | 38.17 | **-1.46** | -0.10 | -0.29 | -0.43 | **-0.83** |
| 6005 | -3.81 | 38.16 | **-1.10** | -0.16 | 0.17 | 0.27 | -0.35 |
| 6011 | -5.96 | 37.98 | **-0.98** | -0.07 | 1.05 | 0.27 | -0.13 |
| 6014 | -6.09 | 37.72 | -0.77 | 0.15 | 0.98 | 0.47 | 0.14 |
| 6020 | -3.68 | 37.27 | **-0.97** | -0.35 | -0.09 | -0.60 | **-0.76** |
| 6022 | -5.76 | 37.04 | **-1.08** | -0.01 | 1.43 | 0.10 | -0.31 |
| 6062 | -4.68 | 38.04 | -0.63 | 0.19 | **1.42** | 0.69 | 0.15 |
|            |           |          | Potential Evapotranspiration (ET$_0$) |        |        |        |        |
|            |           |          | *Winter* | *Spring* | *Summer* | *Autumn* | *Annual* |
| 5012 | -2.67 | 37.89 | **0.32** | 0.15 | **0.12** | -0.06 | **0.16** |
| 6001 | -2.80 | 38.17 | 0.19 | -0.09 | -0.05 | **-0.24** | -0.03 |
| 6005 | -3.81 | 38.16 | **0.21** | 0.10 | 0.06 | -0.08 | 0.06 |
| 6011 | -5.96 | 37.98 | **0.50** | **0.41** | **0.22** | 0.11 | **0.35** |
| 6014 | -6.09 | 37.72 | **0.52** | **0.47** | **0.28** | 0.12 | **0.41** |
| 6020 | -3.68 | 37.27 | **0.28** | 0.11 | -0.01 | -0.02 | 0.06 |
| 6022 | -5.76 | 37.04 | **0.16** | -0.01 | *-0.08* | -0.04 | -0.02 |
| 6062 | -4.68 | 38.04 | **0.23** | 0.08 | -0.05 | **-0.30** | -0.01 |

Table 5. Seasonal and annual trends (in percentage respect the median) of the streamflow, precipitation and potential evapotranspiration in the selected locations. Significant trends at 90% and 95% significance level are in italic and bold, respectively.

Table 6. Changes in land-cover for the stations comprising the "full reference data set". Note that station 5138 does not have any values, as all of its watershed is outside Andalucía, where percentages cannot be calculated from the existing information provided by the Regional Government.

| Code | Name | CN56 | CN77 | CN84 | CN99 | % Change 56-99 | % Surf. with change | % Cultivated land | | % Grassland | | % Woods and forest | | % Others | |
|---|---|---|---|---|---|---|---|---|---|---|---|---|---|---|---|
| | | | | | | | | 1956 | 1999 | 1956 | 1999 | 1956 | 1999 | 1956 | 1999 |
| 5012 | EA-5012 El Doctor | 70.07 | 69.32 | 69.33 | 69.36 | -1.02 | 5.63 | 2.95 | 1.96 | 0.66 | 0.16 | 96.34 | 97.59 | 0.04 | 0.29 |
| 5021 | EA-5021 Quentar | 64.76 | 62.53 | 61.81 | 67.28 | 3.88 | 64.29 | 6.24 | 3.66 | 3.55 | 2.66 | 88.79 | 92.01 | 1.43 | 1.67 |
| 5024 | EA-5024 Hornos del Vidrio | 77.85 | 77.33 | 77.08 | 76.43 | -1.82 | 20.48 | 43.11 | 49.24 | 5.78 | 3.14 | 48.09 | 43.88 | 3.02 | 3.75 |
| 5039 | EA-5039 Central Diechar | 72.66 | 71.58 | 70.66 | 69.05 | -4.97 | 44.74 | 2.45 | 0.10 | 17.84 | 10.81 | 79.25 | 81.09 | 0.46 | 8.00 |
| 5056 | EA-5056 Gerena | 65.77 | 65.88 | 65.80 | 66.67 | 1.37 | 35.18 | 4.08 | 3.70 | 10.63 | 18.34 | 82.84 | 74.35 | 2.45 | 3.61 |
| 5057 | EA-5057 Guadaira | 62.45 | 65.26 | 68.19 | 72.94 | 16.80 | 63.67 | 92.58 | 55.29 | 0.77 | 11.92 | 3.18 | 2.13 | 3.47 | 30.66 |
| 5086 | EA-5086 Dilar | 72.89 | 70.69 | 70.44 | 69.79 | -4.26 | 26.41 | 3.21 | 0.06 | 18.89 | 20.02 | 73.87 | 75.73 | 4.03 | 4.20 |
| 5095 | EA-5095 Tozar | 75.65 | 75.97 | 76.06 | 76.12 | 0.62 | 11.15 | 76.83 | 79.70 | 2.98 | 2.35 | 18.16 | 14.13 | 2.04 | 3.81 |
| 5097 | EA-5097 Los Piedros | 66.02 | 66.21 | 66.20 | 65.89 | -0.20 | 8.47 | 78.10 | 78.90 | 2.58 | 1.86 | 17.20 | 16.29 | 2.11 | 2.95 |
| 5123 | EA-5123 Madrefuentes | 74.04 | 73.94 | 73.84 | 73.97 | -0.09 | 22.20 | 84.33 | 96.62 | 2.70 | 0.32 | 11.10 | 0.90 | 1.87 | 2.17 |
| 5134 | EA-134 Cerrada del Utrero | 66.34 | 65.33 | 65.34 | 64.95 | -2.11 | 15.17 | 0.10 | 0.00 | 5.46 | 5.02 | 93.77 | 93.88 | 0.67 | 1.10 |
| 5137 | EA-5137 Cambil | 74.92 | 74.54 | 74.21 | 74.06 | -1.14 | 16.06 | 30.18 | 28.37 | 7.09 | 9.01 | 60.50 | 59.95 | 2.23 | 2.67 |
| 5138 | EA-5138 Puertollano | - | - | - | - | - | - | - | - | - | - | - | - | - | - |
| 5142 | EA-5142 Salado de Porcuna | 76.71 | 76.85 | 76.87 | 76.92 | 0.27 | 5.92 | 96.82 | 96.70 | 0.61 | 0.34 | 0.87 | 0.81 | 1.70 | 2.15 |
| 5145 | EA-5145 El Pedroso | 63.69 | 63.38 | 63.32 | 63.37 | -0.50 | 13.10 | 20.47 | 20.02 | 13.95 | 11.90 | 63.66 | 65.45 | 1.92 | 2.63 |
| 5152 | EA-5152 Frailes | 75.79 | 75.22 | 75.21 | 74.73 | -1.40 | 15.67 | 33.80 | 33.66 | 28.61 | 14.51 | 34.24 | 48.00 | 3.36 | 3.83 |
| 6001 | E01 Tranco de Beas | 66.20 | 66.20 | 66.02 | 65.07 | -1.71 | 23.80 | 8.67 | 5.20 | 9.95 | 8.22 | 77.76 | 82.27 | 3.63 | 4.31 |
| 6005 | E19 Rumblar | 70.72 | 70.44 | 69.92 | 67.95 | -3.92 | 35.78 | 2.69 | 2.30 | 14.77 | 14.59 | 79.07 | 79.31 | 3.47 | 3.79 |
| 6011 | E57 El Pintado | 64.42 | 63.71 | 63.71 | 63.63 | -1.22 | 17.56 | 21.42 | 20.92 | 15.56 | 17.68 | 60.02 | 58.20 | 3.00 | 3.20 |
| 6012 | E37 Bembézar | 61.90 | 60.35 | 59.64 | 57.92 | -6.43 | 41.68 | 4.98 | 4.69 | 10.66 | 7.34 | 82.67 | 85.68 | 1.69 | 2.30 |
| 6014 | E64 Cala | 68.72 | 68.52 | 68.40 | 67.83 | -1.29 | 22.85 | 3.84 | 3.21 | 15.90 | 14.87 | 77.13 | 78.03 | 3.14 | 3.89 |
| 6016 | E61 Aracena | 70.43 | 70.50 | 70.82 | 69.83 | -0.86 | 23.18 | 11.13 | 9.43 | 13.78 | 12.17 | 72.77 | 73.62 | 2.32 | 4.77 |

Table 6 (cont.).

| Code | Name | CN56 | CN77 | CN84 | CN99 | % Change 56-99 | % Surf. with change | % Cultivated land | | % Grassland | | % Woods and forest | | % Others | |
|---|---|---|---|---|---|---|---|---|---|---|---|---|---|---|---|
| | | | | | | | | 1956 | 1999 | 1956 | 1999 | 1956 | 1999 | 1956 | 1999 |
| 6017 | E39 Retortillo | 58.98 | 59.17 | 59.01 | 58.76 | -0.37 | 22.00 | 8.25 | 7.91 | 11.42 | 10.33 | 78.87 | 78.52 | 1.46 | 3.24 |
| 6018 | E15 Guadalmena | 73.25 | 75.96 | 73.76 | 71.21 | -2.79 | 40.95 | 24.87 | 22.38 | 6.73 | 6.88 | 66.17 | 65.35 | 2.23 | 5.39 |
| 6020 | E45 Cubillas | 72.80 | 72.58 | 72.70 | 72.45 | -0.48 | 18.61 | 57.87 | 62.72 | 2.79 | 1.66 | 37.85 | 33.21 | 1.49 | 2.41 |
| 6022 | E68 Torre del Águila | 76.33 | 76.67 | 76.65 | 76.05 | -0.37 | 12.32 | 81.71 | 86.55 | 4.15 | 2.68 | 10.61 | 7.05 | 3.52 | 3.72 |
| 6029 | E05 Bolera | 70.47 | 69.20 | 69.21 | 68.71 | -2.50 | 20.80 | 3.90 | 2.71 | 7.04 | 6.61 | 88.18 | 88.54 | 0.88 | 2.14 |
| 6037 | E33 Sierra Boyera | 59.56 | 61.22 | 61.32 | 61.03 | 2.47 | 24.58 | 34.91 | 41.38 | 9.76 | 10.45 | 52.86 | 44.49 | 2.47 | 3.68 |
| 6038 | E42 Quéntar | 68.15 | 67.41 | 66.76 | 69.29 | 1.67 | 38.93 | 7.41 | 4.45 | 2.71 | 1.96 | 88.34 | 91.41 | 1.53 | 2.18 |
| 6039 | E17 Quiebrajano | 73.14 | 72.19 | 71.87 | 69.66 | -4.75 | 32.08 | 8.67 | 6.94 | 17.66 | 10.76 | 71.09 | 78.77 | 2.58 | 3.53 |
| 6045 | E02 Aguascebas | 74.88 | 73.13 | 73.05 | 71.76 | -4.17 | 29.21 | 1.90 | 0.98 | 21.52 | 23.27 | 74.15 | 72.07 | 2.42 | 3.69 |
| 6046 | E35 Guadañuno | 56.98 | 57.37 | 57.50 | 57.71 | 1.27 | 20.04 | 0.00 | 0.04 | 22.29 | 18.96 | 76.04 | 76.05 | 1.67 | 4.95 |
| 6047 | E27 Martín Gonzalo | 60.32 | 59.22 | 59.30 | 58.19 | -3.53 | 40.86 | 4.11 | 4.74 | 5.58 | 4.67 | 88.05 | 85.99 | 2.26 | 4.60 |
| 6048 | E41 Canales | 72.99 | 72.35 | 72.21 | 71.38 | -2.20 | 22.41 | 12.49 | 7.75 | 18.74 | 19.75 | 63.52 | 65.75 | 5.26 | 6.75 |
| 6049 | E26 Yeguas | 60.23 | 59.18 | 58.76 | 58.19 | -3.40 | 32.55 | 3.22 | 1.98 | 13.14 | 8.19 | 82.64 | 87.39 | 1.00 | 2.44 |
| 6050 | E44 Colomera | 75.24 | 75.20 | 75.16 | 74.66 | -0.78 | 20.16 | 57.33 | 59.74 | 6.61 | 3.33 | 34.69 | 33.80 | 1.37 | 3.14 |
| 6052 | E14 Fernandina | 72.89 | 71.57 | 70.76 | 69.64 | -4.46 | 44.17 | 4.57 | 2.73 | 20.46 | 19.92 | 72.30 | 72.13 | 2.67 | 5.22 |
| 6055 | E54 José Torán | 61.37 | 61.10 | 61.07 | 61.39 | 0.04 | 17.97 | 21.27 | 21.03 | 16.23 | 14.49 | 60.37 | 59.99 | 2.13 | 4.49 |
| 6056 | E56 Huesna | 64.47 | 64.20 | 64.15 | 64.40 | -0.11 | 15.63 | 18.91 | 18.06 | 16.36 | 15.29 | 62.65 | 62.76 | 2.08 | 3.90 |
| 6058 | E67 Agrio | 64.11 | 60.92 | 61.00 | 60.40 | -5.79 | 55.66 | 0.97 | 0.57 | 7.22 | 5.08 | 88.91 | 89.41 | 2.89 | 4.94 |
| 6060 | E03 San Clemente | 70.55 | 70.08 | 69.98 | 69.92 | -0.89 | 17.54 | 21.66 | 15.30 | 1.88 | 1.31 | 75.82 | 81.58 | 0.64 | 1.81 |
| 6061 | E55 Puebla de Cazalla | 73.97 | 74.21 | 74.23 | 73.97 | -0.01 | 12.52 | 45.18 | 43.62 | 6.82 | 5.93 | 45.91 | 47.61 | 2.09 | 2.83 |
| 6062 | E29 Guadalmellato | 60.76 | 59.89 | 59.88 | 59.75 | -1.67 | 17.40 | 24.70 | 24.08 | 11.94 | 9.53 | 60.54 | 63.29 | 2.82 | 3.10 |

|         | E6001   | E6014  |
|---------|---------|--------|
| Fall    | -0.83   | +0.93  |
| Winter  | **-1.75*** | - 0.47 |
| Spring  | **-1.40*** | **- 1.01** |
| Summer  | **-1.00** | +0.73  |
| Annual  | **-1.59*** | - 0.96 |

Table 7. Seasonal trends (in percentage relative the median) of streamflow simulations in the example basins. Significant trends at 90% significance level are in bold, and at 95% are indicated with an asterisk.

|  | STATION 6001 | | | | | STATION 6014 | | | | |
|---|---|---|---|---|---|---|---|---|---|---|
| $p$ | $\Delta Q/Q_0$ (%) | $S_p$ | $CV_p$ | $CV^*_{Q-p}$ | $CV_Q$ | $\Delta Q/Q_0$ (%) | $S_p$ | $CV_p$ | $CV^*_{Q-p}$ | $CV_Q$ |
| $H_{max}$ | -2.81 | -0.56 | 5.13 | -2.88 | **29.17** | -3.22 | -0.64 | 4.17 | -2.68 | **30.27** |
| $P_{ref}$ | -11.29 | 2.26 | -7.39 | -16.69 | | -11.79 | 2.36 | -7.39 | -17.43 | |
| $ET_0^{ref}$ | -3.51 | -0.70 | 3.75 | -2.63 | | -3.65 | -0.73 | 3.75 | -2.74 | |
| $P_w$ | -2.99 | 0.60 | -38.86 | -23.21 | | -2.96 | 0.59 | -38.86 | -23.00 | |
| $P_f$ | -1.67 | 0.33 | 12.88 | 4.31 | | -3.23 | 0.65 | 12.88 | 8.31 | |
| $P_s$ | 0.69 | -0.14 | -1.01 | 0.14 | | 1.13 | -0.23 | -1.01 | 0.23 | |

\* This is the coefficient of variation of the annual streamflow when the effects of the changes in $H_{max}$ and the forcing series are studied individually

Table 8. Results of the sensitivity analysis.